\documentclass[conference]{IEEEtran}
\IEEEoverridecommandlockouts

\usepackage{cite}
\usepackage{amsmath,amssymb,amsfonts}
\usepackage{algorithmic}
\usepackage{graphicx}
\usepackage{textcomp}
\usepackage{xcolor}

\usepackage{cite}
\usepackage{placeins}
\usepackage{graphicx}
\usepackage{xcolor}
\usepackage{adjustbox}
\usepackage{multirow}
\usepackage{subfigure}
\usepackage{url}

\makeatletter
\newcommand\HUGE{\@setfontsize\Huge{29}{39}}
\newcommand\SMALL{\@setfontsize\small{7}{11}}
\newcommand{\mmio}[1]{{\em memory-mapped I/O}}
\newcommand{\Mmio}[1]{{\em Memory-mapped I/O}}
\newcommand{\comment}[1]{{}}
\newcommand{\commentout}[1]{{}}

\def\BibTeX{{\rm B\kern-.05em{\sc i\kern-.025em b}\kern-.08em
    T\kern-.1667em\lower.7ex\hbox{E}\kern-.125emX}}
\makeatother

\begin{document}

\title{Power and Performance Analysis of Persistent Key-Value Stores}


\author{\IEEEauthorblockN
{\rm Stella Mikrou$^1$, Anastasios Papagiannis$^1$, Giorgos Saloustros, Manolis Marazakis, and Angelos Bilas$^1$}\\
 \textit{Institute of Computer Science (ICS),
Foundation for Research and Technology -- Hellas (FORTH), Greece}\\
100 N. Plastira Av., Vassilika Vouton, Heraklion, GR-70013, Greece \\
Email: \{mikrou, apapag, gesalous, maraz, bilas\}@ics.forth.gr\\
}


\maketitle
\thispagestyle{plain}
\pagestyle{plain}
\footnotetext[1]{Also with the Department of Computer Science,
University of Crete, Greece}
\begin{abstract}
With the current rate of data growth, processing needs are becoming
difficult to fulfill due to CPU power and energy limitations. 
Data serving systems and especially persistent key-value stores have  
become a substantial part of data processing stacks in the data center, 
providing access to massive amounts of data for applications and services. 
Key-value stores exhibit high CPU and I/O overheads because of their 
constant need to reorganize data on the devices. 
\commentout{Today, with the 
introduction of new storage technologies, such as flash (SSDs), 
alternative designs have been proposed that reduce CPU overheads by 
trading device efficiency for CPU utilization.}

In this paper, we examine the efficiency of two key-value stores on
four servers of different generations and with different CPU 
architectures. We use RocksDB, a key-value that is deployed widely, 
e.g. in Facebook, and Kreon, a research key-value store that has been 
designed to reduce CPU overhead. We evaluate their behavior and overheads 
on an ARM-based microserver and three different generations of x86 servers. 
Our findings show that microservers have better power efficiency in the
range of 0.68-3.6x with a comparable tail latency. 
\end{abstract}

\begin{IEEEkeywords} 
data center, energy efficiency,
microservers, key-value stores
\end{IEEEkeywords}

\section{Introduction}
\label{sec:intro}
Projections for data growth show that data doubles roughly every two
years~\cite{dataage} leading to high demand for more processing
capacity to serve and process the data. Given current technology
limitations for power and energy~\cite{kontorinis2012managing, lim2011power}, 
the increasing demand for CPU cycles can only be satisfied by increasing the 
processing density within existing power and energy budgets.
A potential approach is to execute certain classes of
applications on microservers rather than high-end
servers. Microservers include CPUs with different, lower-power
designs, such as ARM processors, compared to typical data center
servers that use higher-end Intel or AMD processors.

Previous research~\cite{ou2012energy, aroca2012towards,
blem2013power, rajovic2013low, Loghin:2015:PSB:2752939.2752945} has
examined the benefits of running applications on microservers 
and in certain cases microservers have been deployed in production
setups~\cite{lai2015atlas, armdatacenters}. 
These previous works use mobile, desktop, web server, database, and 
other workloads to examine performance and energy tradeoffs.

Persistent key-value (KV) stores are a main component of data
analytics stacks and data access frameworks in
general~\cite{decandia2007dynamo, chang2008bigtable, cassandraweb,
  hbase, lai2015atlas, leveldb, rocksdb}.
Typically, persistent KV stores are complex systems because they
constantly re-organize data on storage devices to achieve high data
rates for write, scan, and read operations.  As such, each
user-initiated operation in KV stores requires several thousands of
CPU cycles in the common path~\cite{kreon}.  Recently, new designs for
KV stores have emerged that trade storage device efficiency for CPU
efficiency, in an effort to increase data serving
density~\cite{papagiannis2016tucana, lai2015atlas}.

In this paper, we provide an extensive power/performance 
evaluation for persistent KV stores. This evaluation is important
to calculate the energy budget required for specific performance needs.
We use two KV stores:
RocksDB~\cite{rocksdb}, a persistent KV store from Facebook that is
widely deployed in production setups.  Furthermore, we use
Kreon~\cite{kreon}, a research key-value store that reduces CPU 
overhead and therefore CPU cycles for each KV operation. In order 
to provide a thorough analysis, we use four different,  
server-grade systems that span a broad range of processor 
architectures, memory hierarchy characteristics, and fabrication 
process technologies. We run the default workloads of 
YCSB~\cite{workloads} covering a wide range of cloud use-cases.

We evaluate power efficiency, absolute performance, and architectural
characteristics that affect performance and tail latency. We use one ARM
microserver and three different generations of data center x86
servers. These servers cover a wide range of fabrication processes
technology, micro-architectural features, and amount of processing and
memory resources. For performance experiments we carefully select KV store
configuration setups to make our evaluation realistic.
For power measurements we use a power monitor connected right after the power supply unit (PSU). 
For both performance and power measurements we  perform a large number of experiments 
and we present the most relevant data. Finally, we also provide an analysis for how
different server types contribute to the total cost of ownership in
data centers when used for data serving. Our results show that 
microservers: 
\begin{enumerate}
\item Are 0.68-3.6x times more power efficient.
\item Result in a 1.27-5.3x lower absolute performance where a major factor to performance is 
DRAM throughput.
\item Do not have a big impact to tail latency.
\item Have on average 1.1-2.7x lower energy cost.
\item Are more cost-effective if they have a purchase price around 3x lower than high-end servers.
\end{enumerate}

The rest of the paper is organized as
follows. Section~\ref{sec:related_work} compares our work with related research.
Section~\ref{sec:experimental_methodology} presents our
experimental methodology and Section~\ref{sec:results_and_discussion}
presents our experimental analysis. Finally, Section~\ref{sec:conclusion}
summarizes the conclusions from our experimants.

\section{Related Work}
\label{sec:related_work}

\begin{table*}[t]
\centering
\caption{Our evaluation servers and their hardware
  components.}
\begin{tabular}{|l|r|r|r|r|r|r|r|r|r|r|r|}
\hline
&CPU (all 64-bit)     &\# chips &Fabrication  &ISA &\# cores &\# Threads      &Clock     &L1      &L2 KB     &L3      & DRAM x DIMM\\
&                     &         &technology   &    &         &                &GHz       &KB/core &KB/2core  &MB/chip &GB\\\hline \hline
S1 &X-Gene 1 ARMv8     &1       &40 nm        &ARM &8    &8               &2.40      &32        &256         &8       &16x1=\textbf{16} DDR3 \\\hline
S2 & Xeon(R) E5520     &2       &44 nm        &x86 &8    &16              &2.27      &32        &256         &8       &2x6=\textbf{12} DDR3\\\hline
S3 &Xeon(R)  E5620     &2       &32 nm        &x86 &8    &16              &2.40      &32        &256         &12      &2x12=\textbf{24} DDR3\\\hline
S4 &Xeon(R) E5-2630 v3 &2       &22 nm        &x86 &16   &32              &2.40      &32        &256       &20        &32x8=\textbf{256} DDR4\\\hline
\end{tabular}
\label{table:servers}
\end{table*}

Previous work has compared microservers and high-end servers 
in terms of performance and energy consumption. The authors
in~\cite{Ferdman:2012:CCS:2150976.2150982} show that current 
high-end Out-Of-Order processor micro-architectures are inefficient 
for running scale-out (cloud) workloads. They use performance counters 
to identify key micro-architectural needs for these workloads and 
sources of inefficiency. Also, authors in~\cite{blem2013power} 
revisit the RISC vs. CISC architecture using mobile, desktop, and 
server workloads. They find that RISC and CISC ISAs are irrelevant 
to power and performance characteristics of modern cores, whereas 
micro-architectural features have an important impact on them. In 
our work we use performance monitor counters with the difference that we 
want to identify micro-architectural features that affect the 
performance and power of KV stores. Moreover, we examine how micro-architectural 
features affect tail latency of KV stores.

Also, authors in~\cite{aroca2012towards} and~\cite{ou2012energy} 
compare x86 and ARM architectures in terms of power efficiency 
for web server, database, and other workloads. They conclude that  
x86-based servers are more efficient for compute-intensive workloads 
and ARM-based servers are advantageous in computationally lightweight 
applications in terms of power efficiency. One major difference from 
our work is that we use KV stores and evaluate them on an ARM server 
(not mobile) and on x86 deployed servers. Another difference is that 
we show how micro-architectural features affect performance.  

In~\cite{Andersen:2009:FFA:1629575.1629577}, the authors show that 
low-power embedded nodes with flash storage can deliver over an order 
of magnitude more queries per joule for random read-intensive workloads 
using a custom KV store on a custom cluster. Also in~\cite{lai2015atlas} 
they use a custom KV store that runs to a customized compact server design 
based on ARM processors and they show that is reliable, highly scalable and 
cost-effective. In contrast, in our work we evaluate two persistent key-value 
stores, one from research (Kreon~\cite{kreon}) and one widely deployed in 
production (RocksDB~\cite{rocksdb}), on a range of commodity servers.

Authors in~\cite{Loghin:2015:PSB:2752939.2752945} show that x86 servers 
are more energy efficient in I/O-intensive workloads and ARM servers are 
more energy efficient for database query processing with slightly lower 
performance. They present a total cost of ownership (TCO) analysis and 
they find out that an ARM-based cluster has lower TCO except for I/O 
intensive workloads, where it incurs 50\% higher TCO compared to an 
x86-based cluster. Furthermore, authors in~\cite{ou2012energy} use a monthly 
cost model to analyze cost-efficiency of ARM and x86-based data centers. 
They conclude that from the perspective of cost-efficiency ARM-based data 
centers are advantageous for computationally lightweight applications. 
Our power- and performance-focused analysis can in principle be complemented 
with a similar TCO-focused analysis, to determine the most cost-efficient 
server selection for KV stores.

\commentout{
In our work, we calculate energy cost (dollars per hour) to show the most cost 
efficient server type for KV stores. Furthermore, we include purchase price 
and depreciation time of server and we use a wide range of prices, 3, and 
5 years depreciation time in order to see how these two affect total cost.  
}

\section{Experimental Methodology}
\label{sec:experimental_methodology}

In this section we describe our experimental setups and how we
perform our measurements.

\commentout{\begin{table*}[t]
\centering
\caption{Our evaluation servers and their hardware
  components.}
\begin{tabular}{|l|r|r|r|r|r|r|r|r|r|r|r|}
\hline
&CPU (all 64-bit)     &\# chips &Fabrication  &ISA &\# cores &\# Threads      &Clock     &L1  	  &L2 KB     &L3      & DRAM x DIMM\\
&		      &	    	&technology   &    &	     &		      &GHz	 &KB/core &KB/2core  &MB/chip &GB\\\hline \hline
S1 &X-Gene 1 ARMv8     &1    	&40 nm	      &ARM &8 	 &8 		  &2.40      &32 	&256 	     &8       &16x1=\textbf{16} DDR3 \\\hline
S2 & Xeon(R) E5520     &2    	&44 nm	      &x86 &8 	 &16 		  &2.27      &32 	&256 	     &8       &2x6=\textbf{12} DDR3\\\hline
S3 &Xeon(R)  E5620     &2    	&32 nm 	      &x86 &8 	 &16 		  &2.40      &32 	&256 	     &12      &2x12=\textbf{24} DDR3\\\hline
S4 &Xeon(R) E5-2630 v3 &2    	&22 nm	      &x86 &16 	 &32 		  &2.40      &32 	&256 	   &20 	      &32x8=\textbf{256} DDR4\\\hline
\end{tabular}
\label{table:servers}
\end{table*}
}
\subsection{Server characteristics}
In our work we use four different types of servers that span a broad
range of processor architectures, memory hierarchy characteristics, and
fabrication process technologies. Table~\ref{table:servers} summarizes
the characteristics of each server. Server S1 is an ARM-based server,
whereas servers S2, S3, and S4 are x86 servers of different
generations. All servers have similar clocks (2.27-2.4 GHz), similar
size of L1 cache per core (32KB), and similar L2 cache size (256KB),
shared in all cases by a pair of cores. L3 cache is shared by the
whole chip and the normalized per-core size is similar. Servers S2,
S3, and S4 have two NUMA nodes while S1 has a single NUMA node.  The
number of DIMMs and the total DRAM size differ in all cases, as shown
in Table~\ref{table:servers}.  Finally, servers S1 and S2 use similar
fabrication process technology (40-44 nm), S3 uses a 32 nm process,
and S4 uses a 22 nm process.

We perform various point-to-point comparisons among these servers to
identify characteristics that have an impact on system performance and
power efficiency. To minimize the impact of the software stack, all
servers run the same Linux kernel (version 4.4) with the same version
of the GCC compiler (version 4.8) toolchain. Finally, all servers are
equipped with the same type of NVMe storage device, a Samsung 950 PRO
256GB.



For calibration purposes we measure the memory throughput in each
server using the STREAM~\cite{McCalpin1995} memory benchmark with an
increasing thread count. Table~\ref{tab:memxput} shows these results
for the Triad scenario.

\begin{table}[t]
\centering
\caption{Memory (DRAM) throughput for one thread and a number of
  threads equal to the number cores in the server, as measured with
  STREAM~\cite{McCalpin1995}.}
\begin{tabular}{|l|r|r|r|}\hline
  &\#threads=1 &\#threads=\#cores &\#threads=64\\
  &GB/s        &GB/s   &GB/s\\\hline \hline
S1 &7.7      &8.6  &8.5\\\hline
S2 &9.0      &21.5 &21.2\\\hline
S3 &9.4      &23.9 &23.5\\\hline
S4 &14.2     &68.9 &60.6\\ \hline
\end{tabular}
\label{tab:memxput}
\end{table}	

\subsection{KV stores}
\emph{RocksDB}~\cite{rocksdb} is an LSM-based persistent key-value
store that is widely used in production at Facebook. Is is optimized
for fast storage but it can be also used for hard disk drives. It 
contains multiple levels of increasing size where keys are sorted 
within each level. Each level is divided in multiple units of fixed 
size named SSTable (SST) and each of them is stored in a separate
file.

In RocksDB, the first level is stored in memory (named memtable) and 
when it becomes full it is flushed in first levels SSTs. In order to 
provide write amortization, in keep the ratio between 2 successive 
levels to be less that 10x. In the case where the ratio exceeds 
this value, a compaction is triggered. A compactions merges SSTs of level
$L_{i}$ with SSTs if the next level ($L_{I+1}$). RocksDB interacts with 
storage when it writes a memtable to the device, creating a new SST and 
during compactions. In both cases it uses read-write API to access files.

\emph{Kreon}~\cite{kreon} is a persistent write-optimized key-value
store designed for flash storage. The main design tradeoff is that it
increases I/O randomness in order to reduce CPU overhead and I/O
amplification. To achieve this, Kreon uses a write-optimized data
structure and \mmio{}.

Kreon uses a multi-level indexing data structure similar to the 
LSM-Tree~\cite{o1996log}, with levels of increasing size.
This enables batched data transfers to lower levels to amortize 
insert costs. Kreon uses a per-level full index (B-tree)
to enable partial data reorganization which reduces I/O amplification 
and CPU overhead. Furthermore, Kreon stores key-value pairs in an 
append-only log to avoid data movement during spill (merge) operations.
The use of \mmio{} further reduces CPU overheads related to the I/O
cache in three ways: (a) It eliminates cache lookups for hits by using
valid virtual memory mappings. Accesses to data that are not in memory
result in page faults that are then handled by Linux kernel. (b)
\textit{Read}/\textit{write} system calls require a data copy between
user and kernel space for protection purposes. \Mmio{} removes the
need for data copies when performing I/O. (c) \Mmio{} uses a single
address space for both memory and storage, which eliminates the need
for pointer translation between them. Therefore, this approach removes
the need for serialization and deserialization when transferring data
between memory and storage.

\subsection{Workloads}
We run a C++ version of YCSB~\cite{Jinglei2016, workloads} using the
proposed workloads in the recommended sequence: Load the database, run
workloads in order A, B, C, F, and D, clear the database, load the
database again and run workload E. In all cases the key is about 30
bytes and the value is 1000 bytes. We use datasets that fit in main
memory for all servers, since we are mainly interested in CPU
efficiency.

For our analysis, we run two different experiments for both key-value
stores. For Kreon the first experiment uses a single YCSB thread and a
single Kreon table with a dataset of 3M records that fits in memory
and does not cause I/O traffic (no snapshots) for all servers. We run 
this experiment to identify the performance and energy characteristics 
of the core's architecture. In the second experiment for Kreon we run 
YCSB with multiple threads to identify server performance and power 
consumption under high utilization. In this case, the duration of each 
run for 3M records is short, which does not allow us to observe server 
performance under steady state. For this reason, and since servers have 
different main memory sizes, we use for each server a dataset proportional 
to its memory. We use 350K records per 1GB of DRAM i.e., 2.4M for S1, 
4.2M for S2, 8.4M for S3, and 89.6M for S4. Finally, we use 64 YCSB 
threads and 32 Kreon tables in all cases to keep all servers at high 
CPU utilization.

For RocksDB the first experiment uses a single YCSB thread and one
database with dataset of 6M records. We enable direct I/O with 2GB block cache. In
this experiment we identify the performance and energy characteristics
of the core's architecture with I/O traffic. In the second experiment
for RocksDB we run YCSB with multiple threads again with direct I/O.
To avoid thread contention and keep utilization high we choose to keep
the number of YCSB threads equal to the number of hardware threads for
all servers (S1 does not support hardware threads and therefore, this
is the number of cores). We use a number of databases equal to half
the number of YCSB threads.

Since our interest is CPU behavior, in our experiments we try to
maintain similar I/O behavior across servers. For this purpose, we
configure the datasets in a manner where the LSM multi-level structure
exhibits the same I/O behavior across servers: We use 4 databases and
5M records for S1, 8 databases and 10M records for S2/S3, and 16
databases and 20M records for S4. With these parameters, the number of flushes
and compactions for each workload are the same for each
server. Finally, we adjust the size of block cache, to fit the same amount of
dataset for each server i.e, 2GB for S1, 4GB for S2/S3 and 8GB for
S4.

\begin{table}[t] 
\begin{center}
\caption{Workloads used with YCSB. All workloads use Zipf
  distribution except for D that uses latest distribution.}
\begin{tabular}{lc} 
  &\textbf{Workload}\\ \hline 
  A &  50\% reads, 50\% updates \\ 
  B &  95\% reads, 5\% updates \\ 
  C &  100\% reads \\ 
  D &  95\% reads, 5\% inserts \\ 
  E &  95\% scans, 5\% inserts \\ 
  F &  50\% reads, 50\% read-modify-write \\ \hline 
\end{tabular} 
\label{tab:wload}
\end{center} 
\end{table}

\subsection{Power measurements}
We measure power for all platforms with a Microchip MCP39F511A Power
Monitor Demonstration Board~\cite{MCP39F511A}. The Power Monitor is connected right after the power
supply unit (PSU), which converts the AC wall socket supply to DC. It
reports mean power readings every second.  This measures the total
power of the board, including CPU, memory, and storage devices.
We use the Linux kernel's {\textit perf} tool to collect and analyze performance
monitor counters (PMC) that are available in all servers.  We
calculate branch miss ratio, instructions per cycle (IPC), and
Last-Level-Cache (LLC) (i.e. L3) miss ratio.  We perform our
calculation based on reading from the following PMCs: branches, branch
misses, instructions, cycles, L3 cache references and L3 cache
misses. The ARM-based server S1 has limited set of performance
counters and it is does not provide counters for the LLC
(L3).

\section{Experimental Analysis}
\label{sec:results_and_discussion}

In this section we focus our analysis around four main questions:
\begin{itemize}
\item Which server architecture is more power efficient?
\item Which server architecture achieves the highest absolute throughput?
\item Which micro-architectural features (do not) matter?
\item Does server performance translate to tail latency benefits?
\end{itemize}
Next, we discuss each of these in detail.

\subsection{Which server architecture is more power efficient?}
\begin{figure}[t]
\centering
\subfigure[Kreon, many threads]{\includegraphics[width=.5\linewidth]{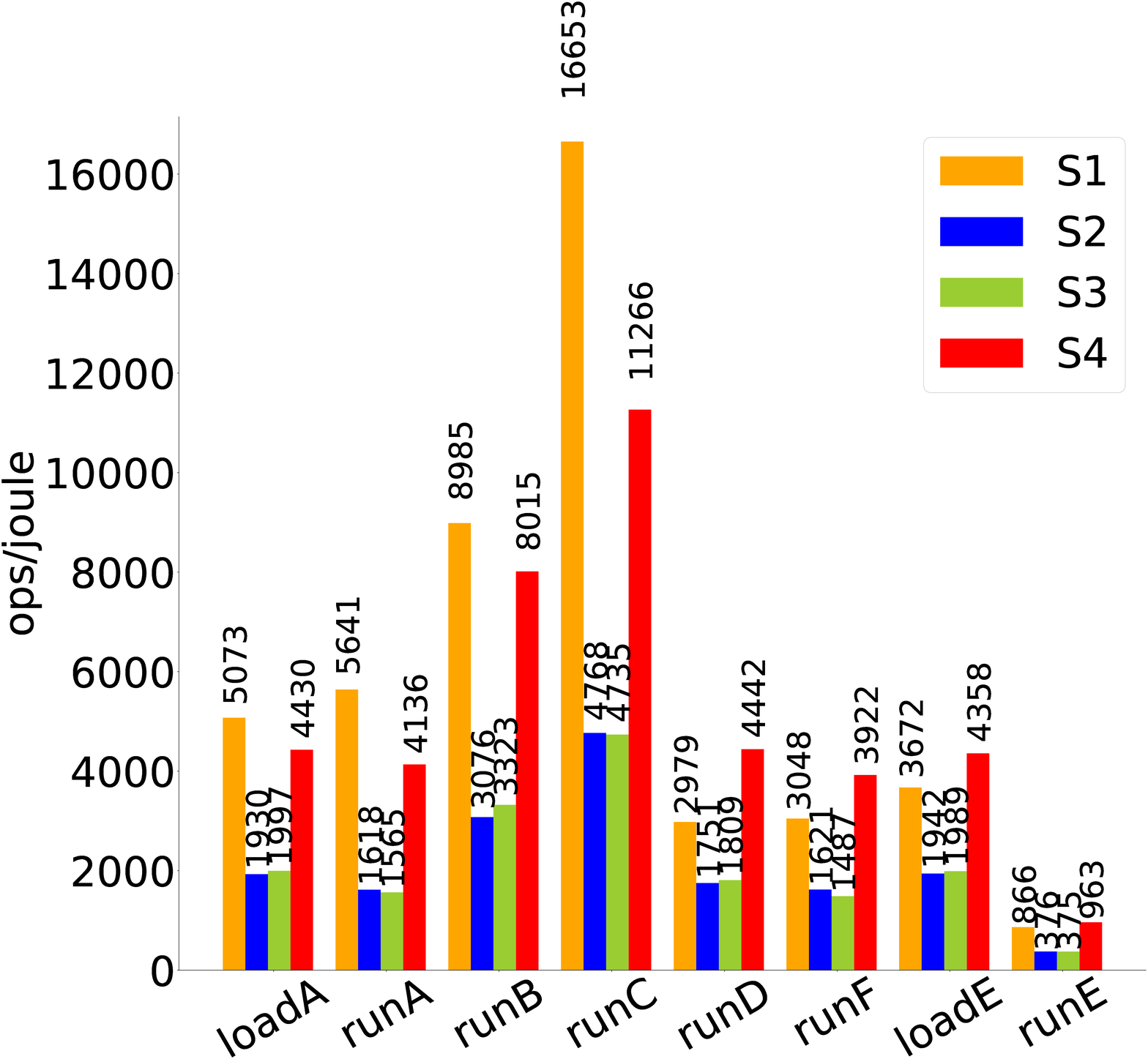}\label{subfig:power_joule}}~\subfigure[Kreon, 1-thread]{\includegraphics[width=.5\linewidth]{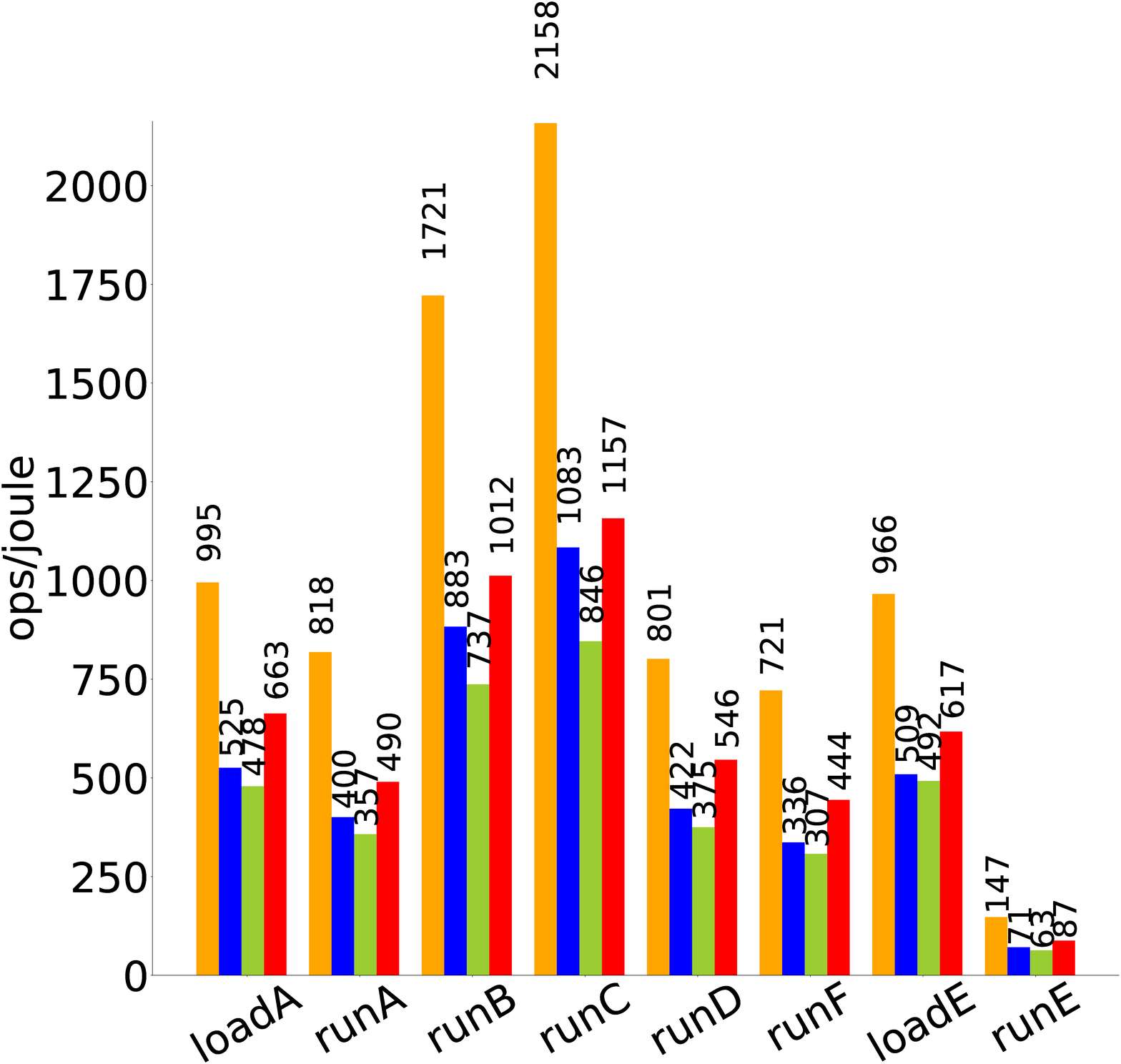}\label{subfig:power_single_thread}}
\subfigure[RocksDB, many threads]{\includegraphics[width=.5\linewidth]{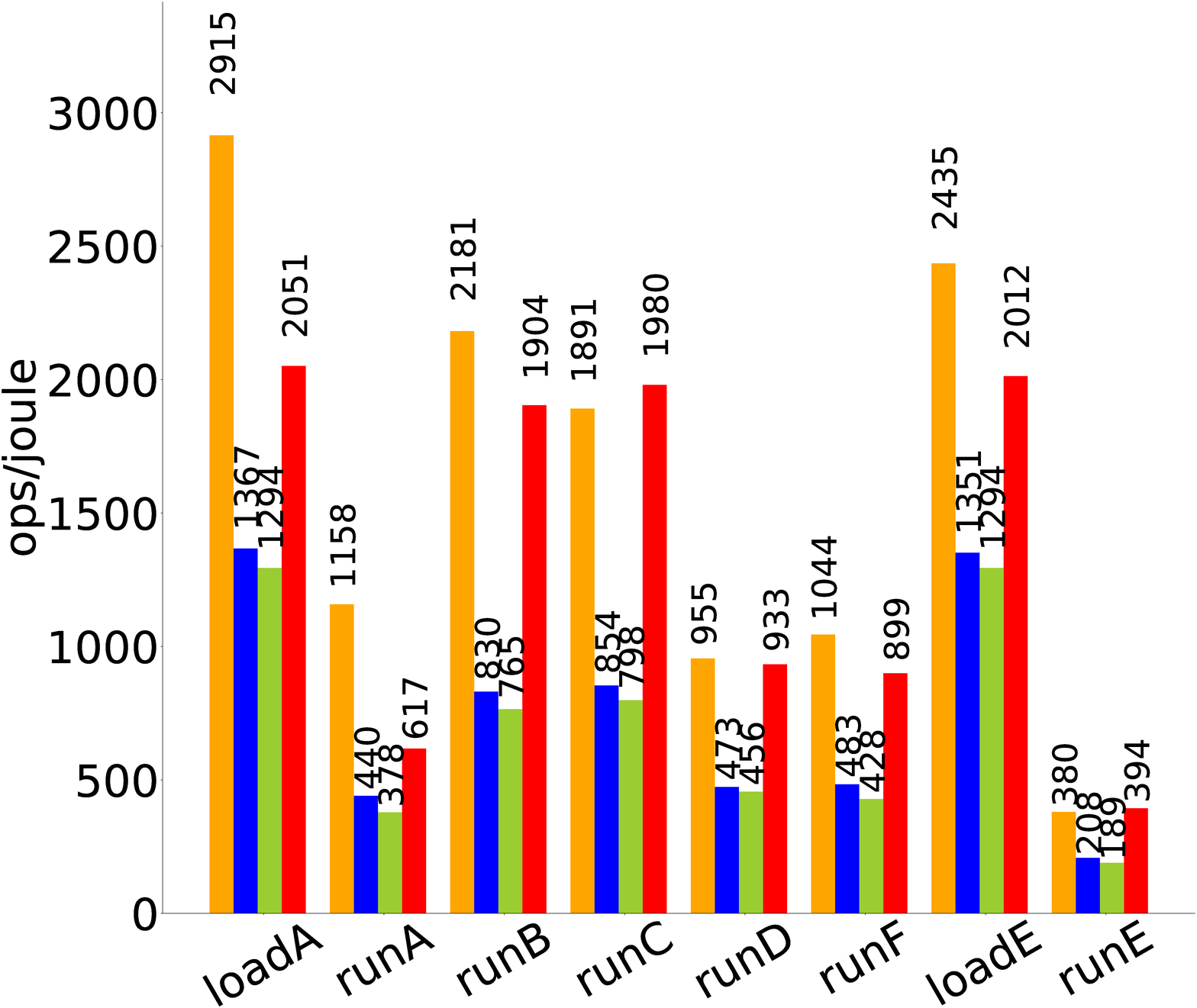}\label{subfig:power_joule_rdb}}~\subfigure[RocksDB, 1-thread]{\includegraphics[width=.5\linewidth]{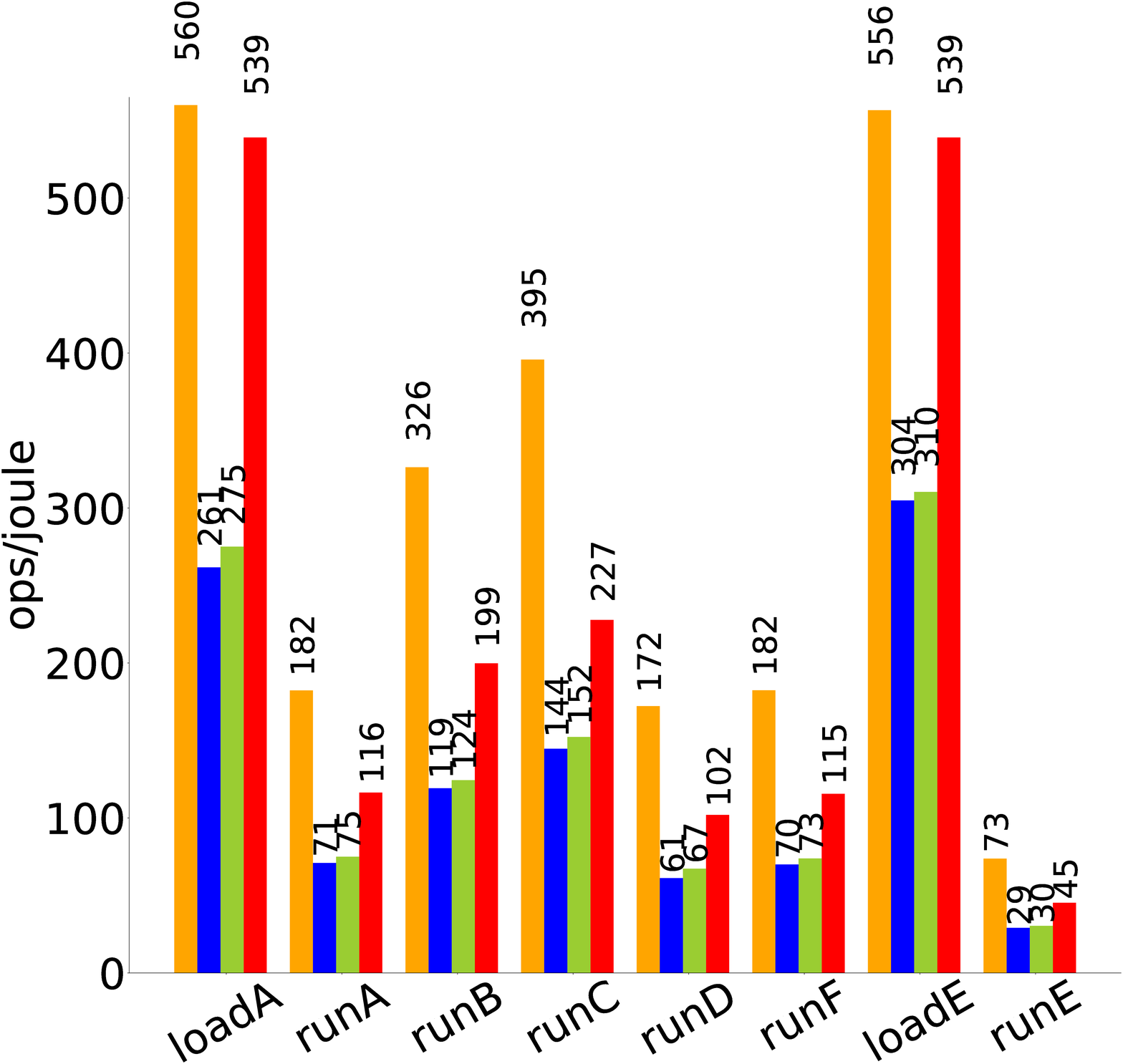}\label{subfig:power_single_thread_rdb}}
\caption{Power efficiency (ops/joule) for Kreon (top) and RocksDB (bottom).}
\label{fig:power}
\end{figure}

We measure power efficiency as ops/joule in all servers for both
KV stores (Figure~\ref{fig:power}).
Figures~\ref{subfig:power_joule} and~\ref{subfig:power_joule_rdb} show
the efficiency of each server for Kreon and RocksDB under high
utilization, above 80\% in all runs shown.  First, for both KV stores,
we can categorize servers in two groups: S2, S3 and S1, S4.  We see
that S1 and S4 are more power efficient than S2 and S3 for both KV
stores.  Compared to servers S2 and S3, S1 executes 1.6-3.6x more
ops/joule for Kreon and 1.8-3x more ops/joule for RocksDB.  Similarly,
compared to S2 and S3, S4 achieves 2.1-2.7x more ops/joule for Kreon
and 1.4-2.5x more ops/joule for RocksDB. We note that servers S2, S3
have the same architecture as S4, but differ significantly in
fabrication process technology (Table~\ref{table:servers}).

Between S1 and S4 there is greater variance. S4 has both a more
aggressive architecture and more recent fabrication process (22nm for
S4 vs. 40nm for S1). However, S1 achieves between 0.68-1.47x more
ops/joule for Kreon and between 0.96-1.87x for RocksDB. We notice 
that S1 is better at writes (Load A, Load E) for both KV stores
except Load E at Kreon that is slightly worse. Finally, S4  
achieves always slightly more operations per joule in scans 
(Run E).

Figures ~\ref{subfig:power_single_thread}
and~\ref{subfig:power_single_thread_rdb} depict ops/joule for the
single-threaded experiment (one YCSB thread and one database) for both
KV stores.  This experiment shows the behavior of a single thread with
abundant resources, including shared micro-architectural resources,
memory throughput, caches, I/O.
We notice that a single thread in server S1 is more power efficient
compared to a single thread in server S4 by 1.46-1.86x for Kreon and
by 1.03-1.74x for RocksDB.

Compared to S2 and S3, S1 exhibits single-thread efficiency
(ops/joule) between 1.89-2.55x for Kreon and 1.79-2.81x for RocksDB.
Compared to S2 and S3, S4 achieves a single-thread efficiency
(ops/joule) between 1.06-1.45x for Kreon and 1.49-2.06x for RocksDB.
Therefore, despite its older fabrication process, S1 is more power efficient
compared to S2 and S3, than S4.

Finally, between servers S2 and S3 we do not observe any significant
differences in ops/joule in both the high-utilization and
single-thread experiments. Their CPU architectures are similar, they 
both have 2 NUMA nodes and they have about the same fabrication process.

Overall, server S1 is better by 0.68-3.6x in terms of ops/joule in all
our experiments for both KV stores, despite the fewer resources and
older fabrication technology.

\subsection {Which server architecture achieves the highest absolute throughput?}
\begin{figure}[t]
\centering
\subfigure[Kreon, kops/s]{\includegraphics[width=0.25\textwidth]{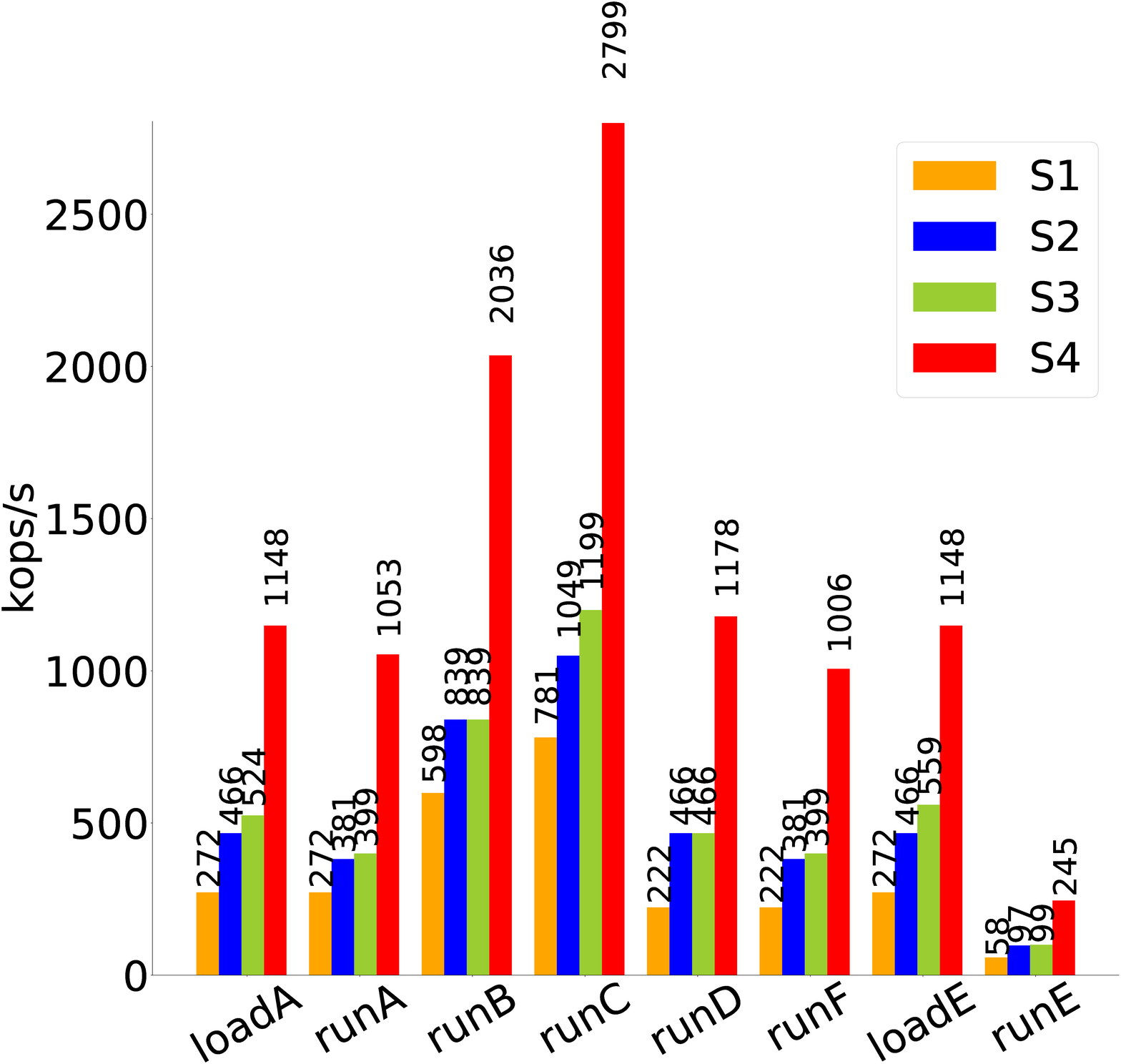}\label{subfig:ops_sec}}~\subfigure[Kreon, IPC]{\includegraphics[width=.25\textwidth]{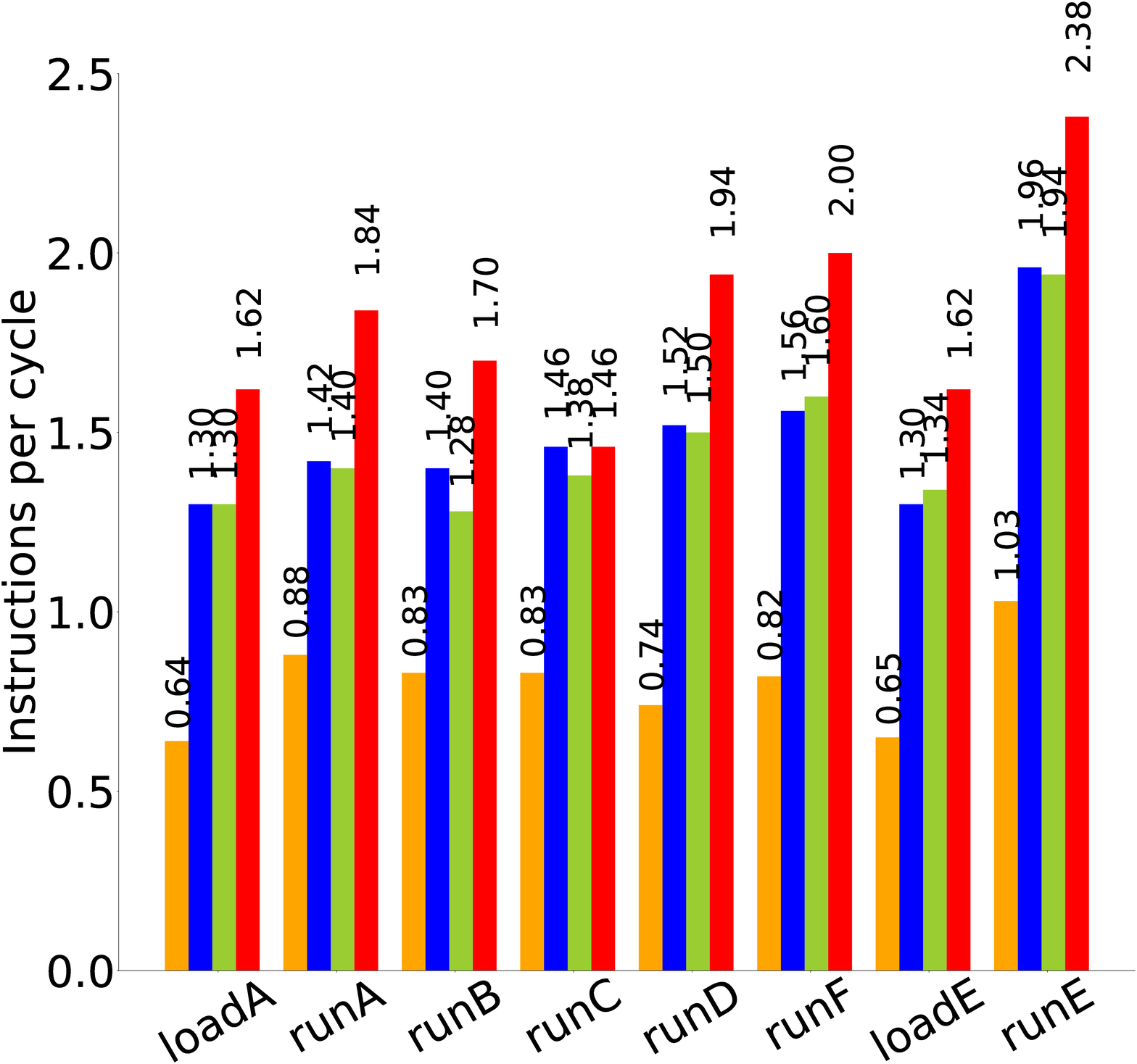}\label{subfig:ipc}}
\subfigure[RocksDB, kops/s]{\includegraphics[width=0.25\textwidth]{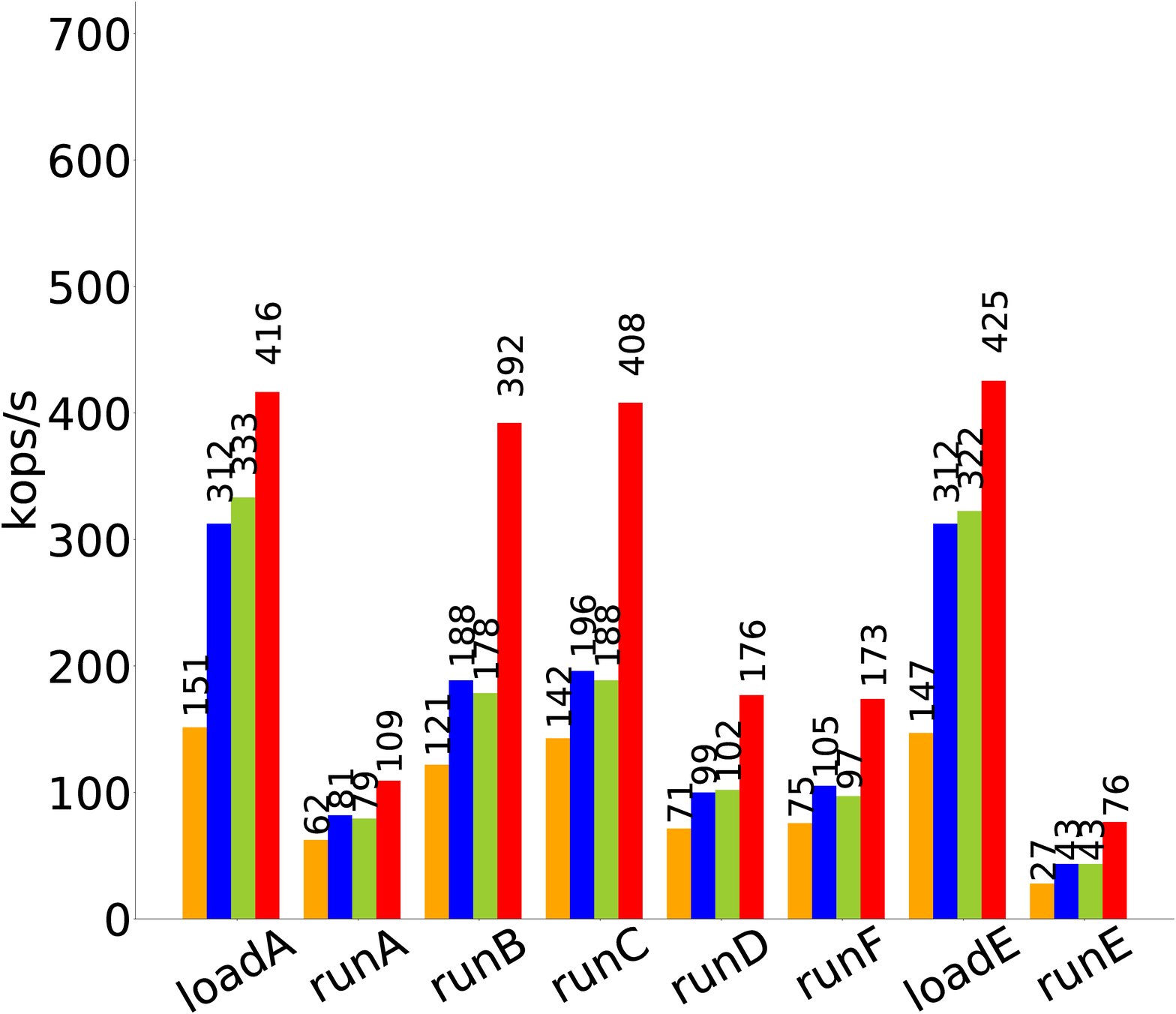}\label{subfig:ops_sec_rdb}}~\subfigure[RocksDB, IPC]{\includegraphics[width=.25\textwidth]{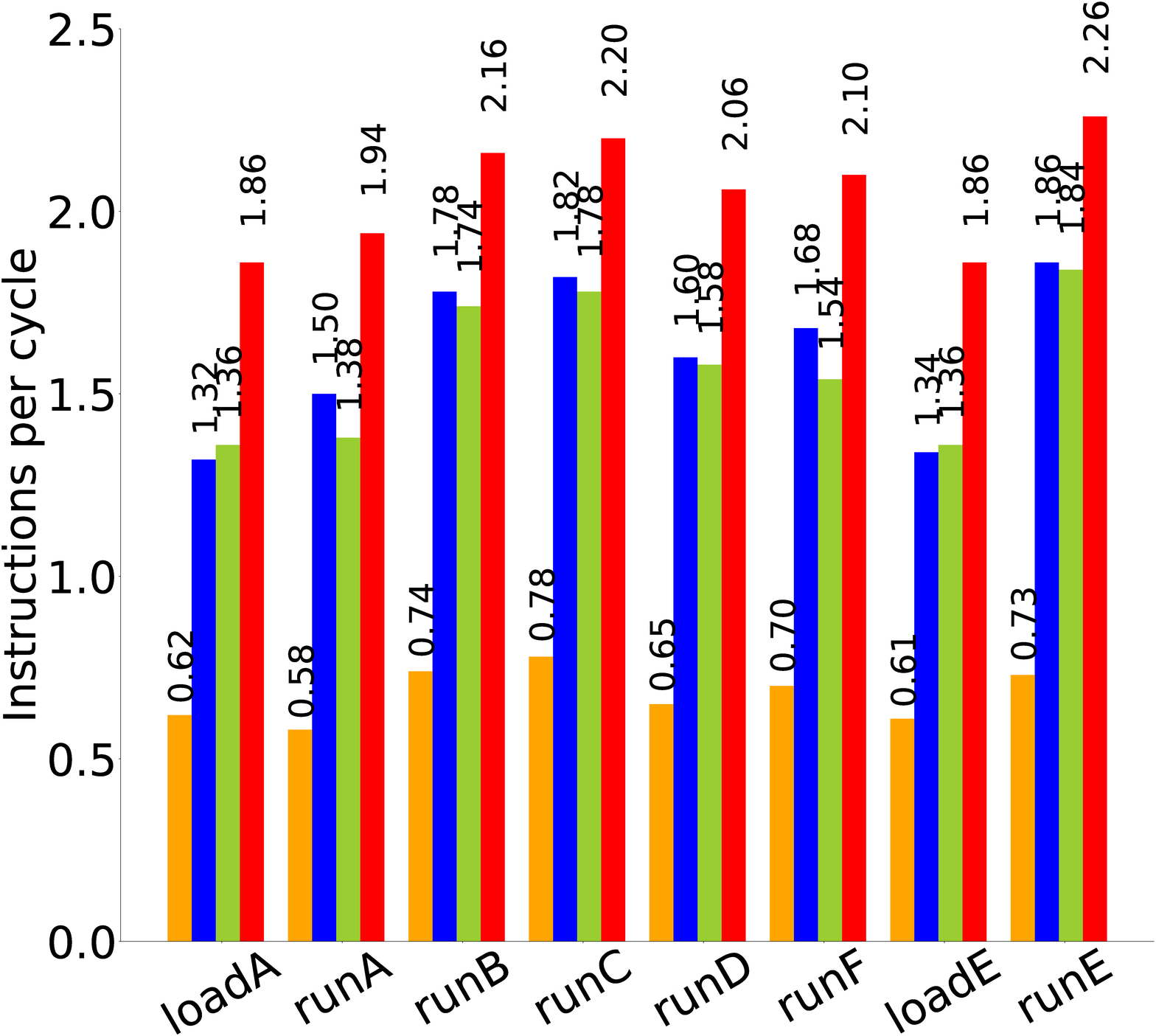}\label{subfig:ipc_rdb}}
\caption{Multi-threaded absolute performance and IPC for Kreon (top)
  and RocksDB (bottom), under high CPU utilization.}
\label{fig:throughput}
\end{figure}

Figures ~\ref{subfig:ops_sec} and ~\ref{subfig:ops_sec_rdb} show
absolute performance expressed in kops/s. We see that servers of
different generations exhibit significant differences in performance
for KV stores up to 5.3x. In Kreon, S1 exhibits up to 5.3x worse
performance (kops/s) compared to S4 and between 1.34-2.0x worse
performance compared to S2 and S3.  Servers S2 and S3 have
approximately the same absolute performance and 2.0-2.7x worse
compared to S4.  In RocksDB, S1 exhibits 1.75-3.23x fewer kops/s
compared to S4 and 1.27-2.2x lower performance compared to S2 and
S3. S4 achieves 1.24-2.2x higher performance compared to S2 and S3.

Next, we examine the achieved IPC per core (not per hardware
thread). Figure~\ref{subfig:ipc} shows that IPC follows the
same trend as absolute performance across all servers. S4 achieves the
highest IPC among all servers in the range of 1.46-2.38, whereas S1
achieves the lowest IPC in the range of 0.64-1.03. If we multiply IPC
with the number of cores in each server, we approximately get the same
trend as absolute performance.

\begin{figure}[t]
\centering
\subfigure[Kreon, kops/s]{\includegraphics[width=0.25\textwidth]{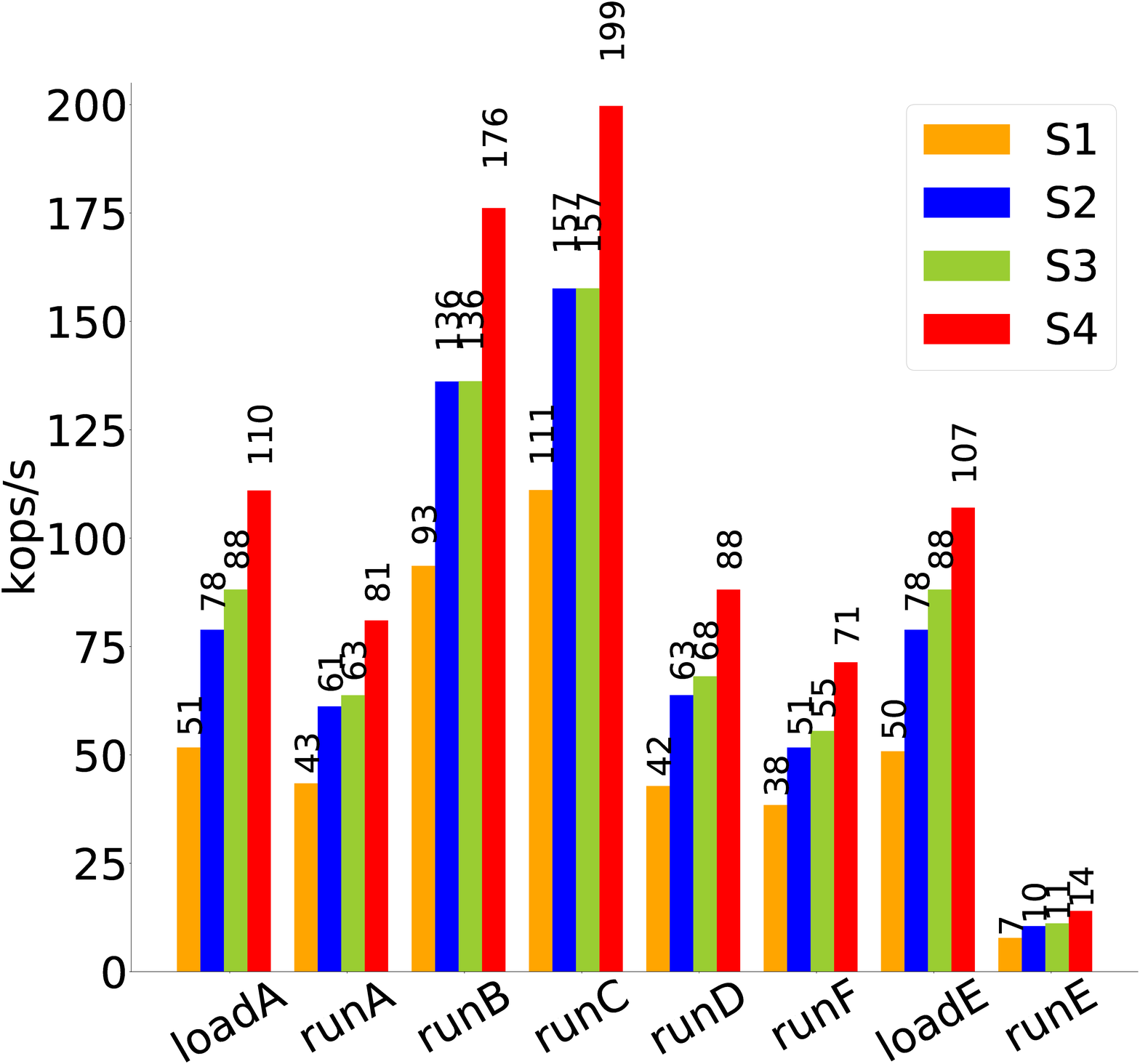}\label{subfig:ops_sec_single}}~\subfigure[Kreon, IPC]{\includegraphics[width=.25\textwidth]{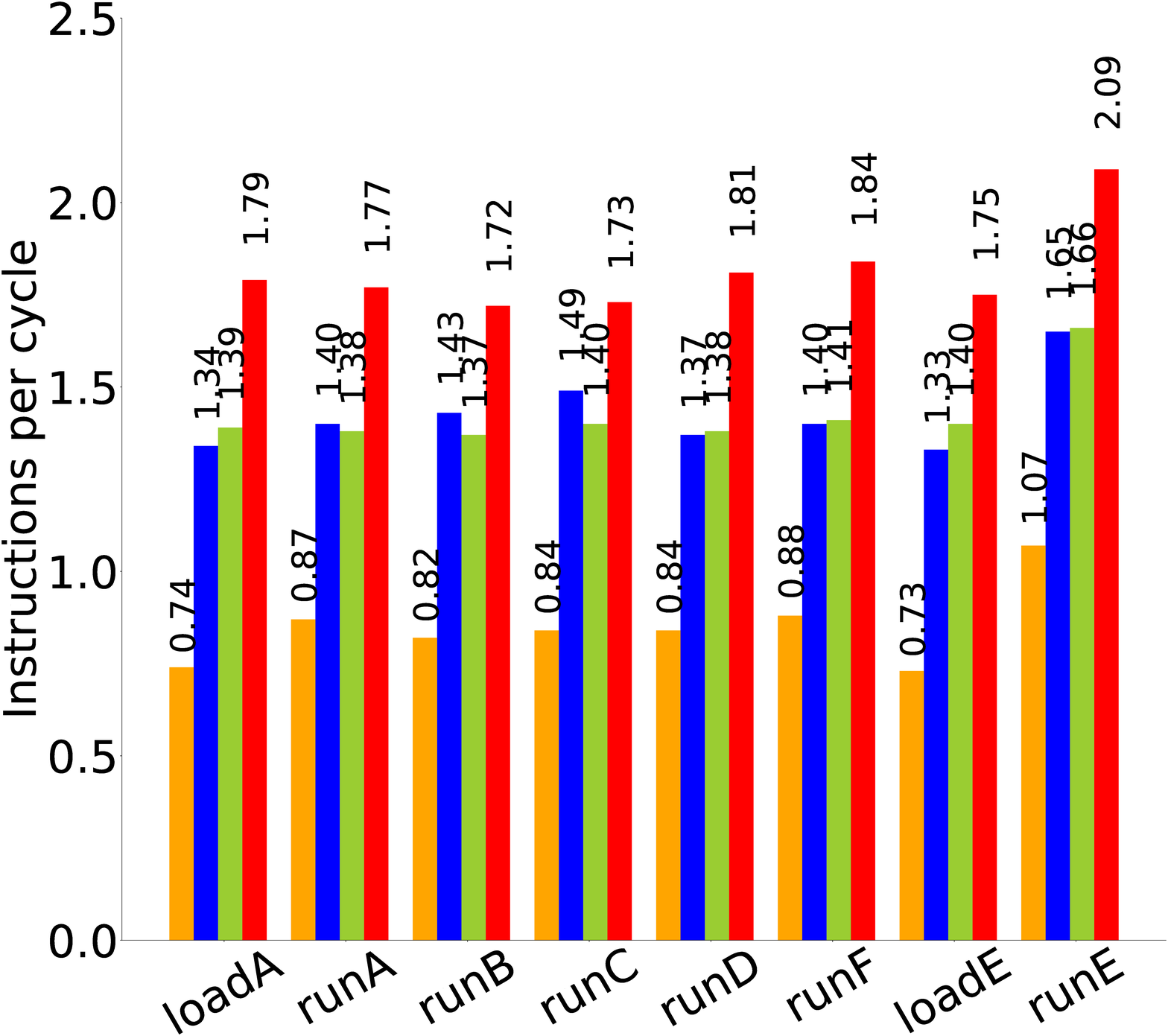}\label{fig:ipc_single}}
\subfigure[RocksDB, kops/s]{\includegraphics[width=0.25\textwidth]{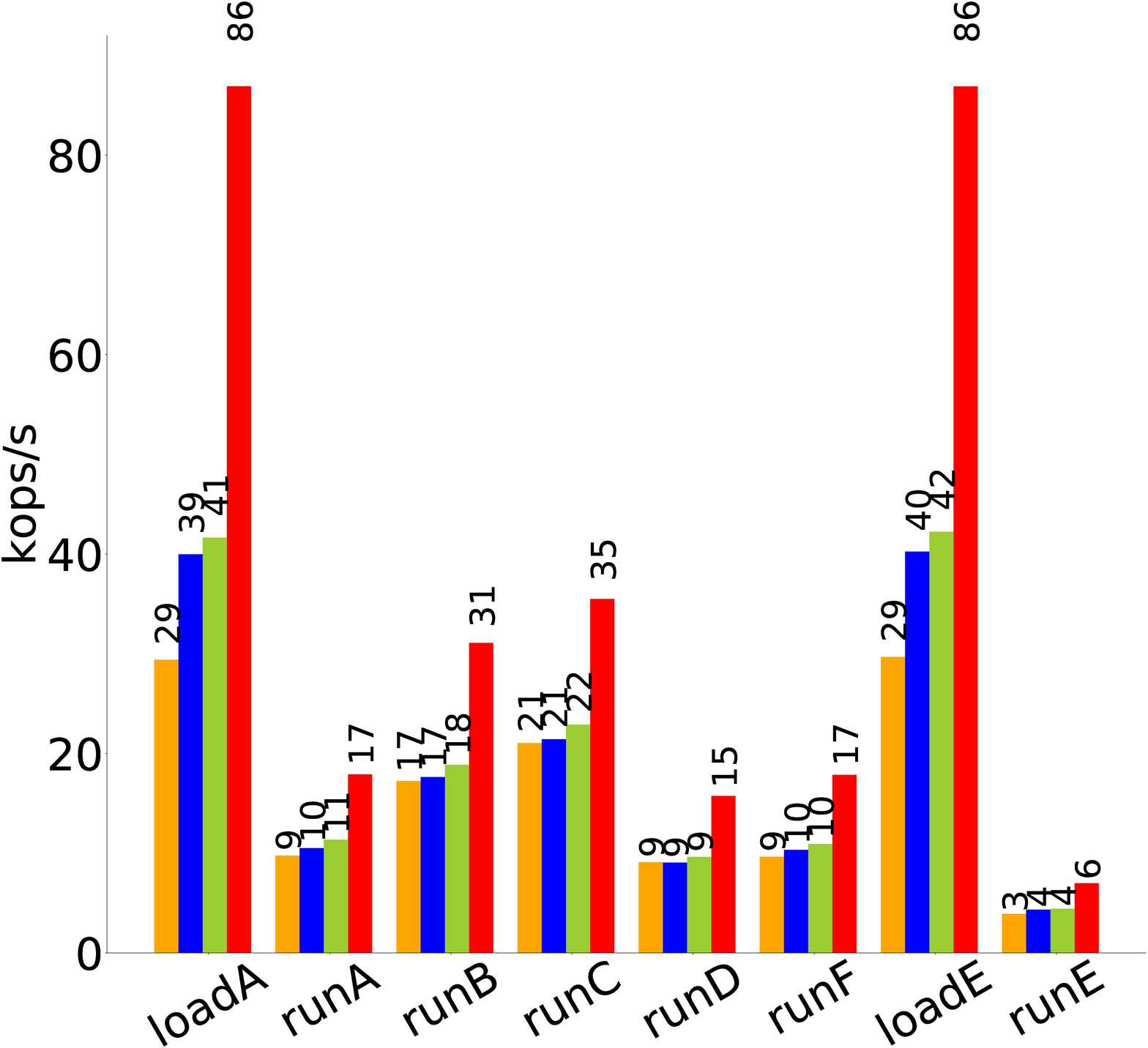}\label{subfig:ops_sec_single_rdb}}~\subfigure[RocksDB, IPC]{\includegraphics[width=.25\textwidth]{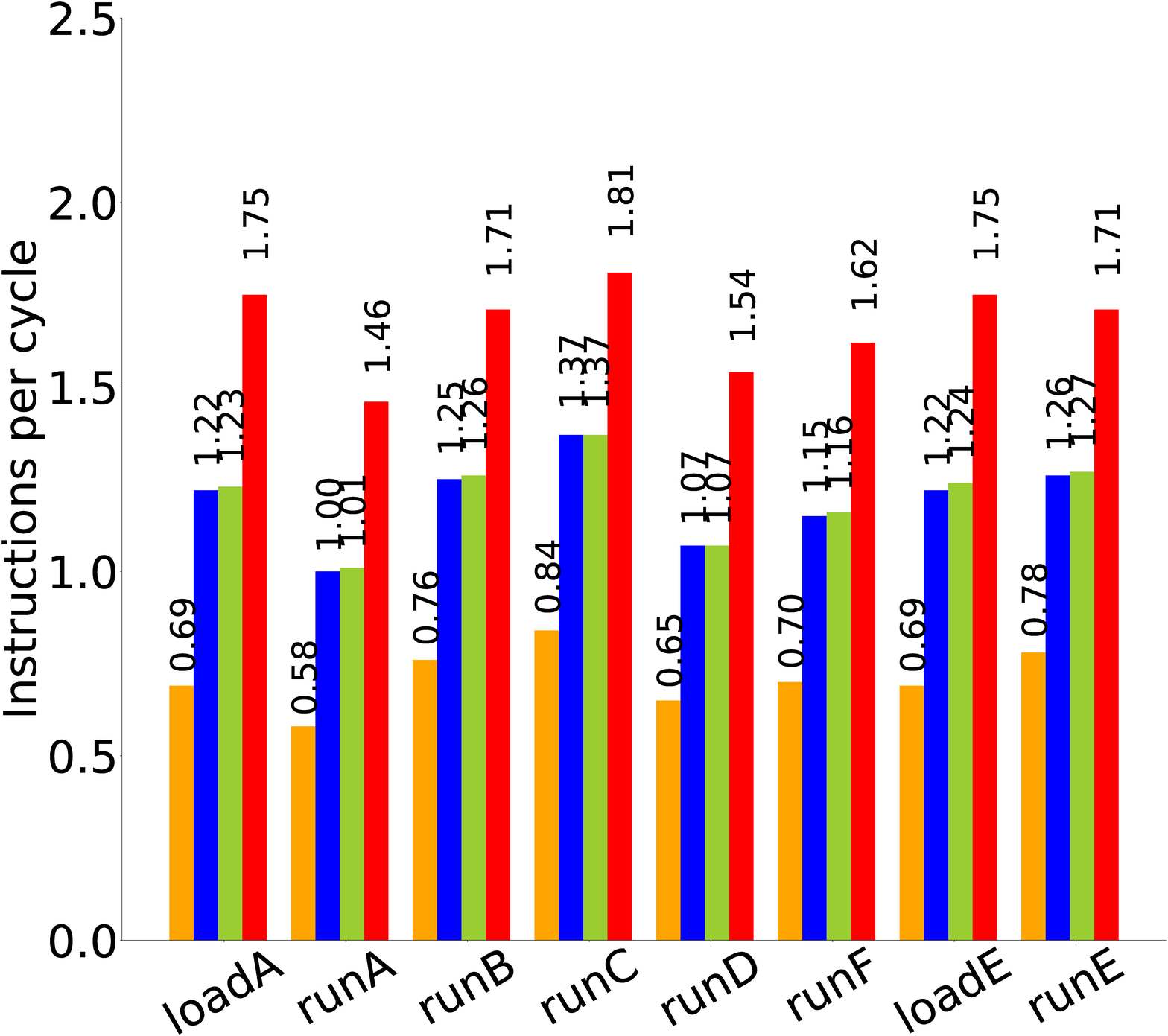}\label{fig:ipc_single_rdb}}
\caption{Single-thread absolute performance and IPC for Kreon (top)
  and RocksDB (bottom).}
\end{figure}

Figure~\ref{subfig:ops_sec_single} captures single thread performance
for each server type. It measures absolute performance (ops/s) with a
single YCSB thread assigned to one core.  We see that a single thread
in S1 performs 1.34-1.76x worse than S2, S3. Compared to S4, S1 has
1.9-2.1x lower throughput. Finally, S4 compared to S2, S3, achieves
1.3-1.4x higher throughput. This experiment, with all
resources available to one thread, shows (Figure~\ref{fig:ipc_single})
that S1 achieves again the lowest IPC, between 0.73-1.07. Server S4
achieves the highest IPC between 1.72-2.09. We get the same trend
across servers, as single thread performance. In comparison in
multi-threaded experiments, the differences in absolute performance and
IPC between servers decreases, compared to single-thread experiment.

\subsection{Which micro-architectural features (do not) matter?}
\begin{figure}[t]
\centering
\subfigure[Kreon branch miss ratio]{\includegraphics[width=.25\textwidth]{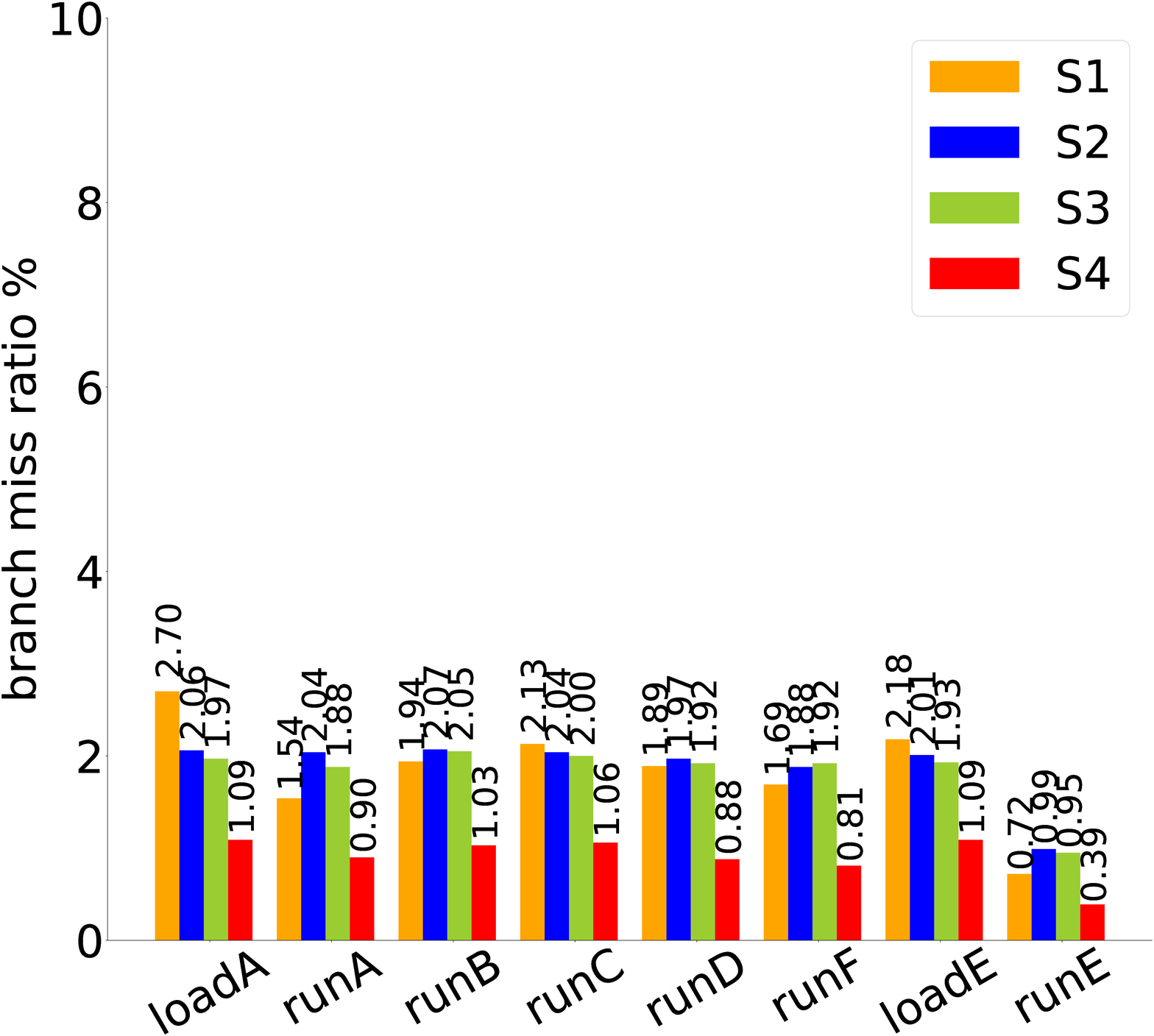}\label{subfig:branch}}~\subfigure[Kreon L3 miss ratio]{\includegraphics[width=.25\textwidth]{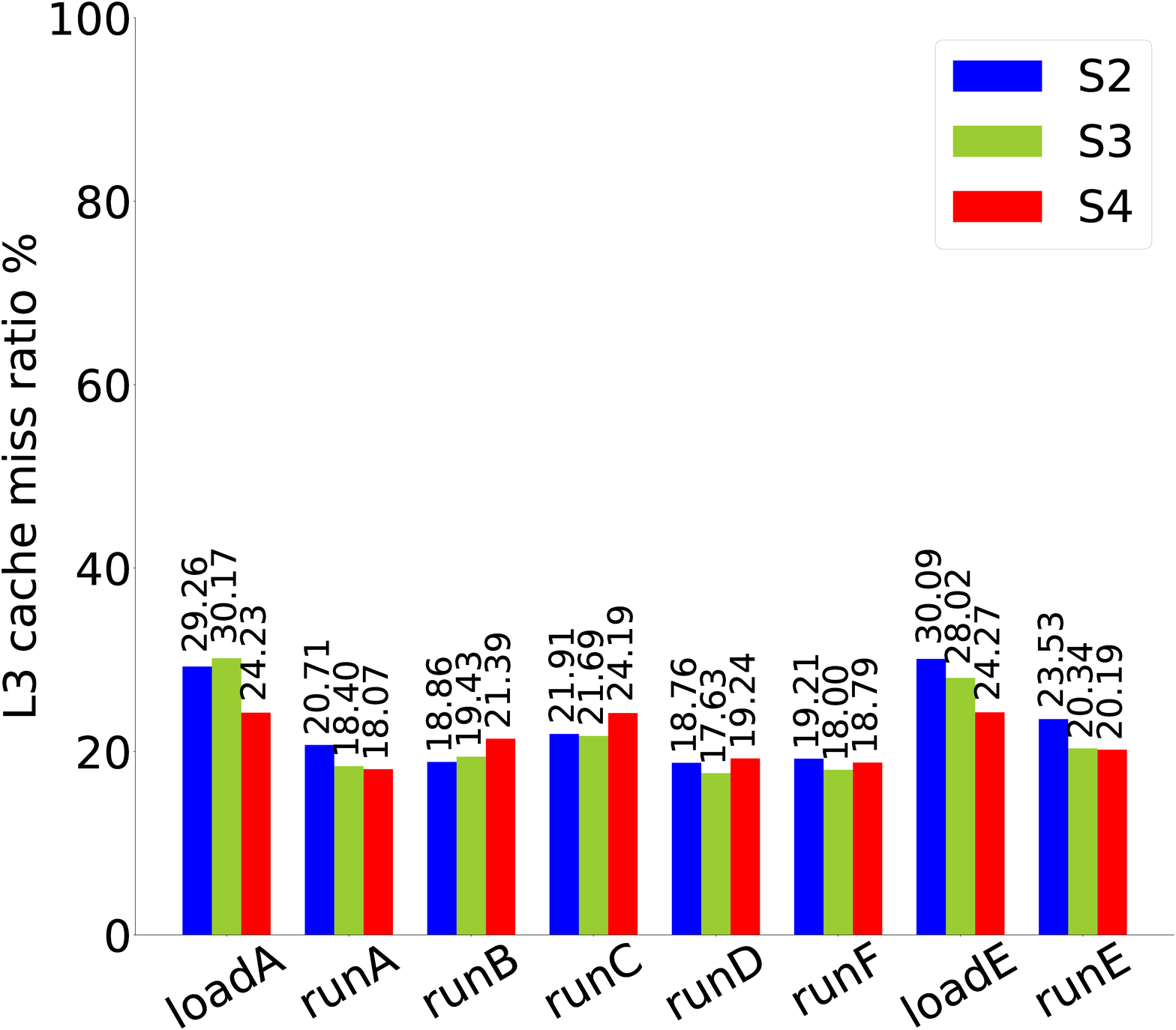}\label{subfig:l3_misses}}
\subfigure[RocksDB branch miss ratio]{\includegraphics[width=.25\textwidth]{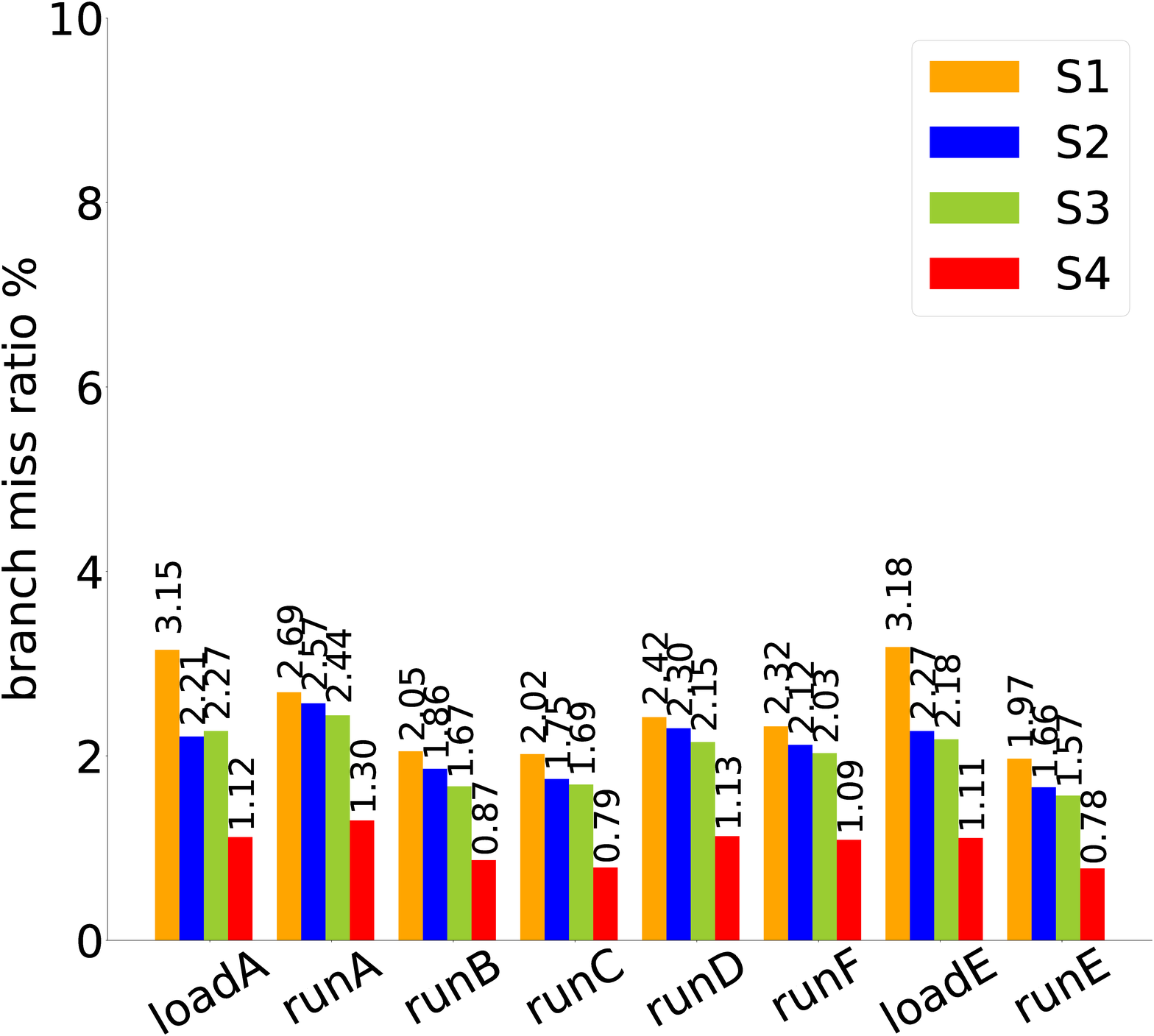}\label{subfig:branch_rdb}}~\subfigure[RocksDB L3 miss ratio]{\includegraphics[width=.25\textwidth]{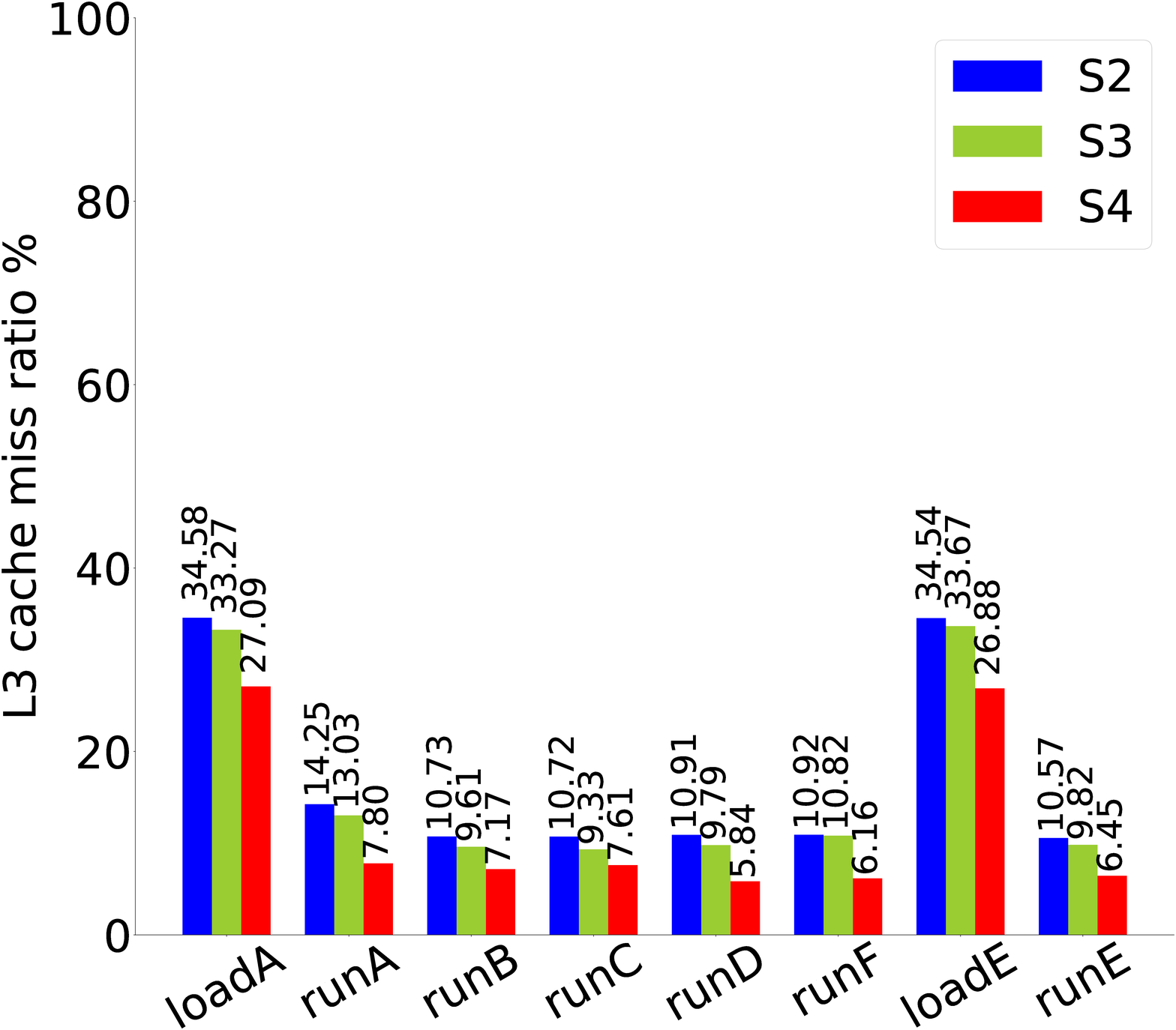}\label{subfig:l3_misses_rdb}}
\caption{Performance monitor counter measurements for branch and L3 cache miss ratio in multi-threaded experiment for Kreon (top) and RocksDB (bottom).}
\label{fig:perf}
\end{figure}

Next, we examine CPU performance counters for several events to
identify sources of performance differences.  We study branch and L3
cache miss ratios, and the impact of hardware multi-threading, as
shown in Figure~\ref{fig:perf}. We perform these measurements for both
multi-threaded and single-threaded experiments for each server. 
Results are averages across all cores.

\paragraph{Branch misses}
Figures~\ref{subfig:branch} and~\ref{subfig:branch_rdb} show that 
the branch miss ratio does not exceed 3.18\% for all servers and workloads 
for both KV stores. We observe that S4 has significantly lower 
branch miss ratio compared to the other servers and in most cases it 
incurs less than 50\% of the misses. However, given the overall low 
branch miss ratio, this does not contribute significantly to the observed 
performance differences.

\paragraph{L3 misses}
As a note, L3 miss ratio is not available for S1 because of counter
limitations on the specific platform. Figures~\ref{subfig:l3_misses}
and~\ref{subfig:l3_misses_rdb} show that the L3 miss ratio differs
between 1-6\% of L3 references across servers, for both Kreon and
RocksDB. Although total L3 cache sizes differ across servers, the 
amount of L3 cache per core is about the same: S1 and S2 have 1 
MB/core, whereas S3 and S4 have 1.5 and 1.25 MB/core, respectively, 
resulting in similar L3 miss ratios.
\begin{figure}[t]
\centering
\subfigure[Kreon]{\includegraphics[width=.25\textwidth]{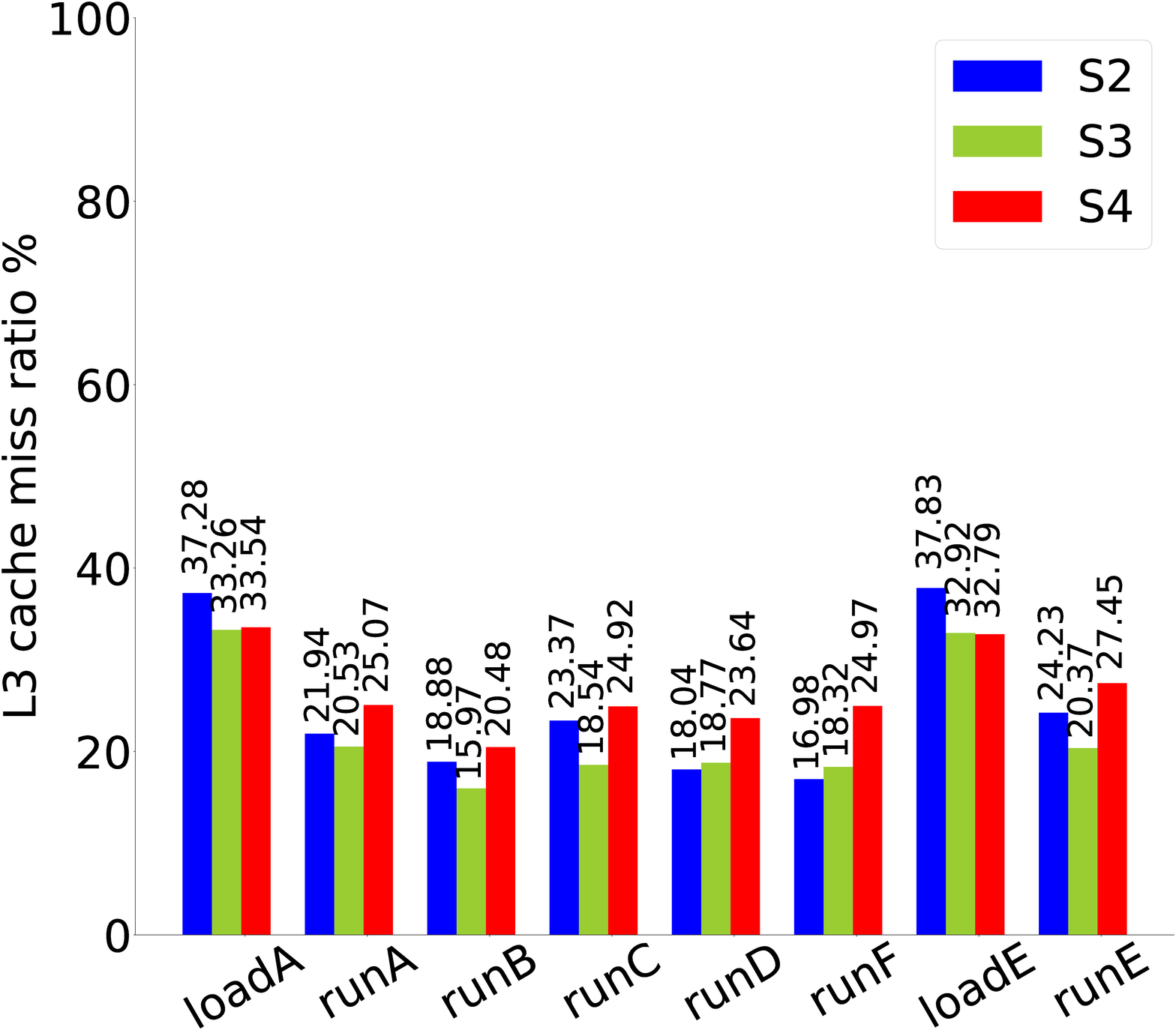}\label{fig:l3_single}}~\subfigure[RocksDB]{\includegraphics[width=.25\textwidth]{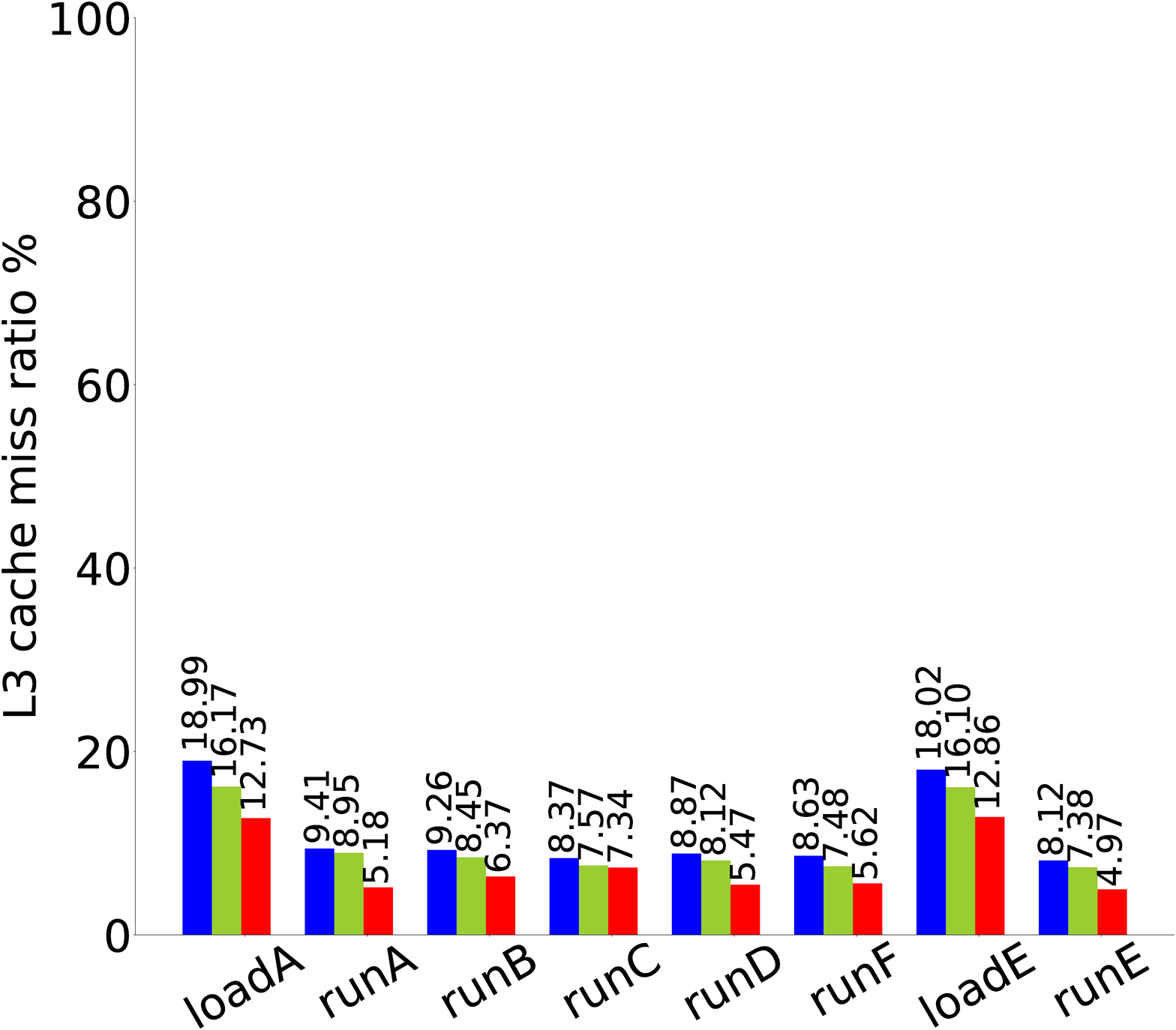}\label{fig:l3_single_rdb}}
\caption{Single-thread L3 miss ratio for Kreon and RocksDB.}
\end{figure}

To examine how larger L3 caches and other components contribute to
performance, we run the single-threaded experiment, with a single YCSB
thread assigned to one core. Figure~\ref{fig:l3_single} shows that the
L3 cache miss ratio differs up to 6\% of all L3 cache references among
all servers for both KV stores. This difference is similar to 
multi-threaded runs (Figures~\ref{subfig:l3_misses} and 
~\ref{subfig:l3_misses_rdb}), although in this case the per core L3 
size differs significantly across servers. Therefore, L3 cache size 
does not contribute significantly to performance.

\paragraph{Hardware multi-threading}
S1 supports a single hardware thread per-core while S2, S3, and S4
have hyper-threading and thus they provide two hardware threads
per-core. We perform the experiment of Figure~\ref{subfig:ops_sec} 
with hyper-threading disabled. We find that servers S2 and S3 
perform 1.22-1.31x fewer kops/s, compared to the same experiment with
hyper-threading enabled, whereas S4 performs 1.15-1.22x fewer kops/s.
Similarly, IPC with hyper-threading disabled, has a drop of 1.22-1.34x 
and 1.16-1.29x for servers S2/S3 and S4 respectively. This shows that 
using twice the number of hardware threads (hyper-threading) only increases 
performance between 1.15-1.34x across all cases.
\begin{figure}[t]
\centering
\subfigure[Kreon]{\includegraphics[width=.5\textwidth]{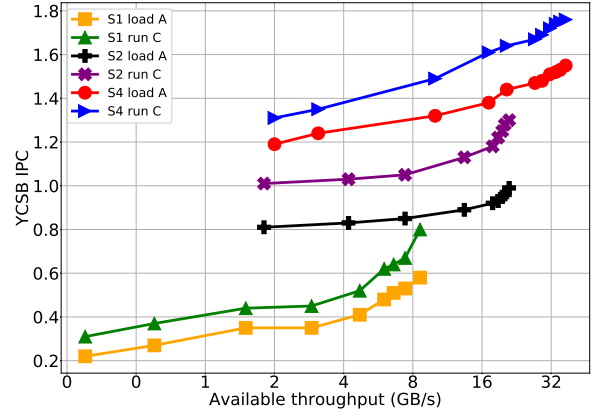}\label{fig:stream_ipc_kreon}}
\subfigure[RocksDB]{\includegraphics[width=.5\textwidth]{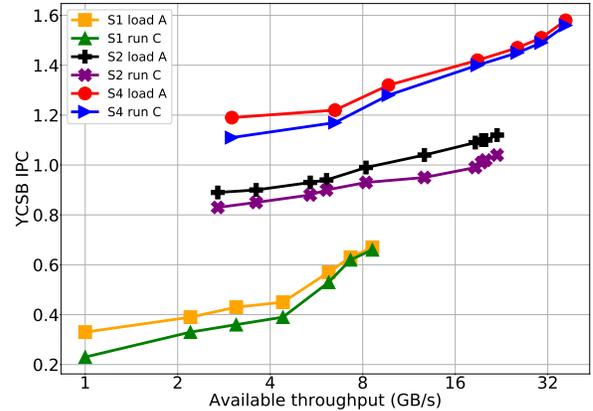}\label{fig:stream_ipc_rdb}}
\caption{IPC for four hardware threads (running YCSB) with an
  increasing available memory throughput.}
\label{fig:stream_ipc}
\end{figure}

\paragraph{Memory throughput} 
DRAM throughput affects IPC. Tbale~\ref{tab:memxput} shows that S1 has 2.5x lower maximum memory throughput 
than S2, S3 and 7x lower maximum memory throughput than S4. The 
differences are smaller for the maximum memory throughput observed 
by a single thread.

To examine how memory throughput affects KV store performance,
we create one microbencmark which can consume a specific amount of 
memory throughput. We run, for both KV stores, four 
YCSB threads concurrently with four threads of our microbenchmark, 
each of them pinned to one physical core. We choose four threads 
because this is the minimum number of threads that can consume the total 
memory throughput. In every run we decrease the throughput consumed by our 
microbenchmark to increase the available throughput for 
Kreon and RocksDB. Figure~\ref{fig:stream_ipc} shows the resulting 
IPC for four YCSB threads running Load A and Run C.
In all cases we achieve better IPC when available 
memory throughput increases. Overall, all systems can benefit 
from higher memory throughput than the currently provisioned 
2.1, 5.4, 9.1 GB/s/core for S1, S2, and S4 respectively.
%

\subsection {Does server performance translate to tail latency benefits?}
\begin{figure}[t]
\centering
\subfigure[Kreon, Load A]{\includegraphics[width=.49\linewidth]{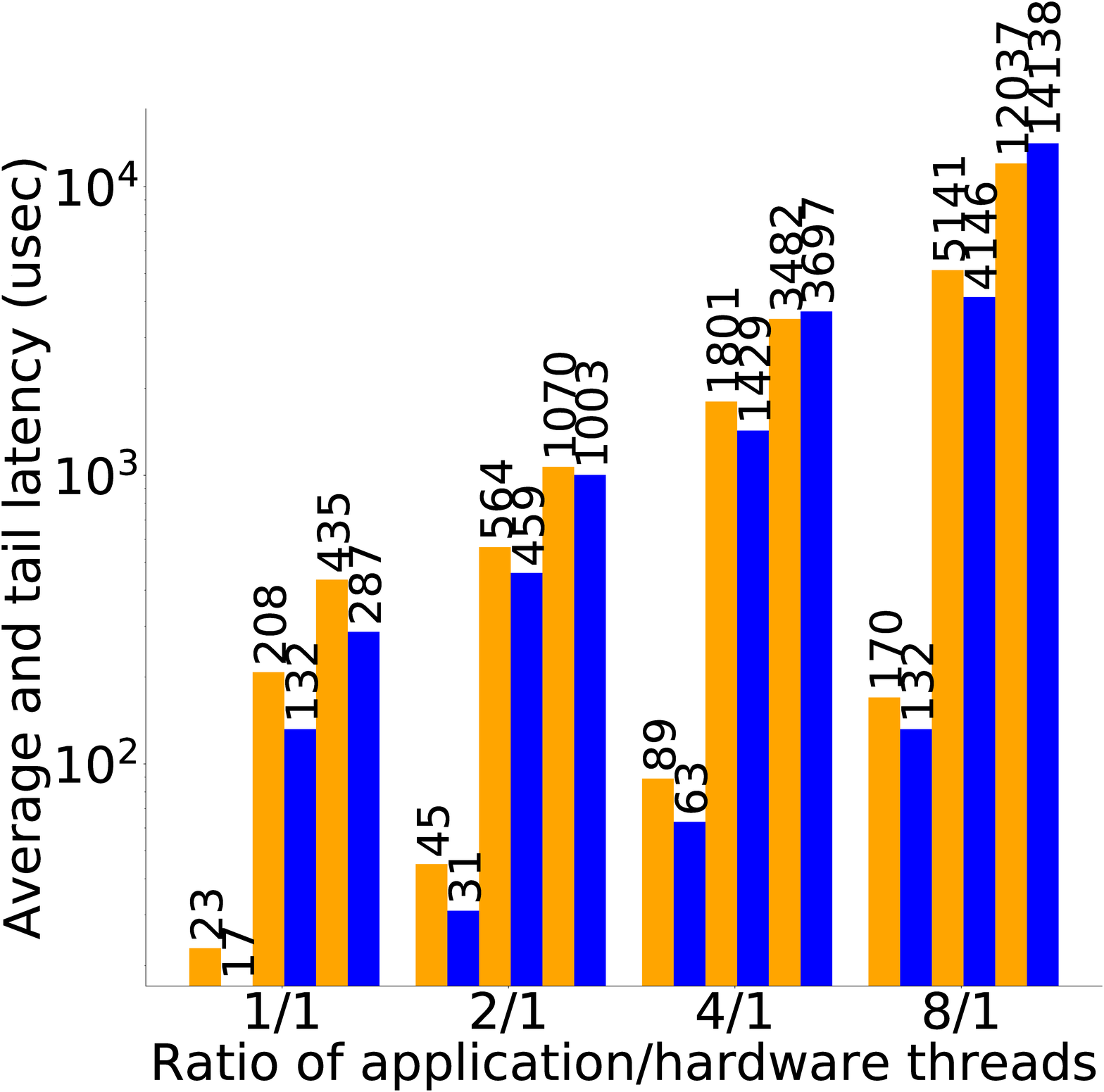}\label{subfig:load_a_avg}}~\subfigure[Kreon, Run C]{\includegraphics[width=.49\linewidth]{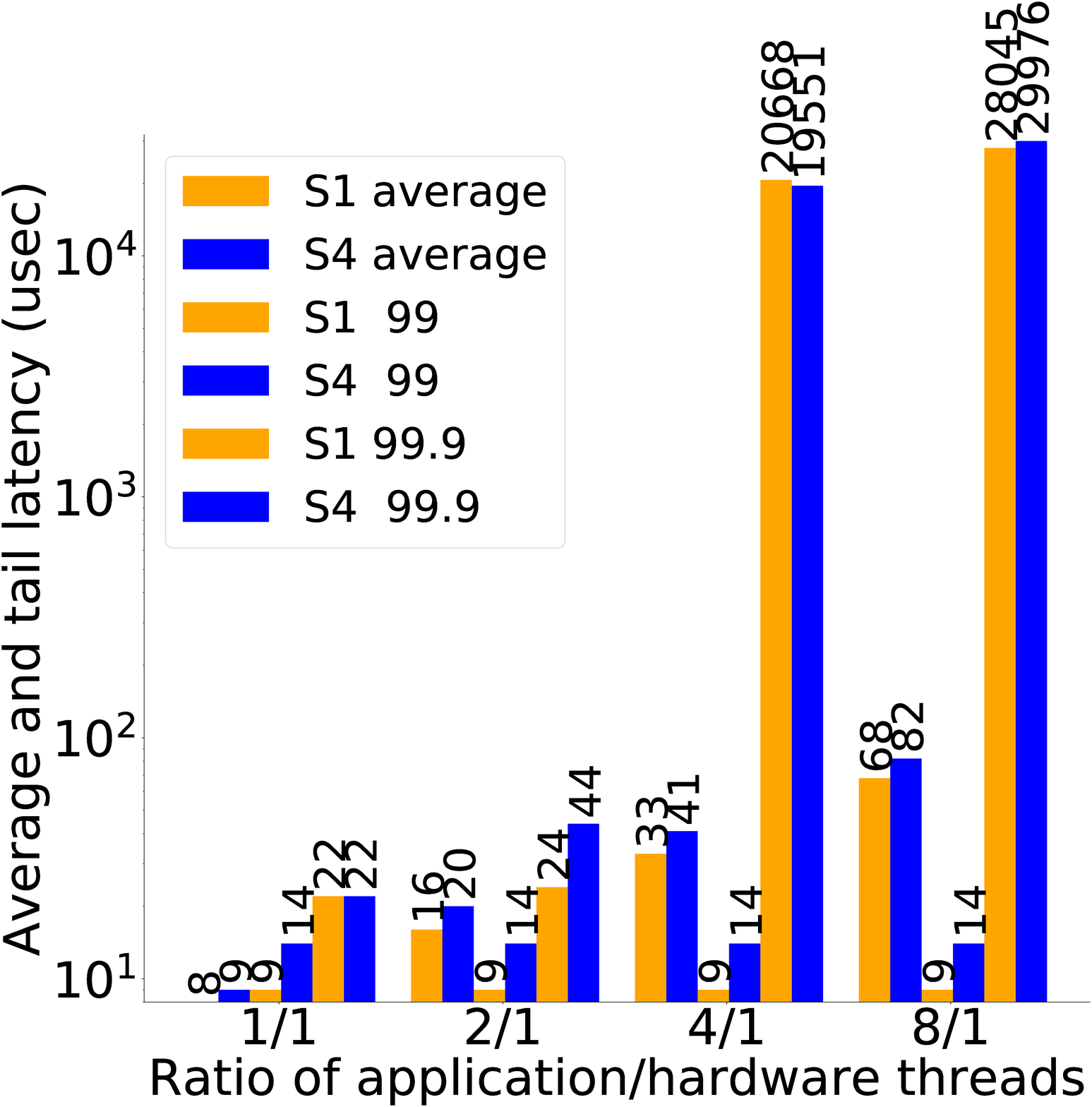}\label{subfig:run_c_avg}}
\subfigure[RocksDB, Load A]{\includegraphics[width=.49\linewidth]{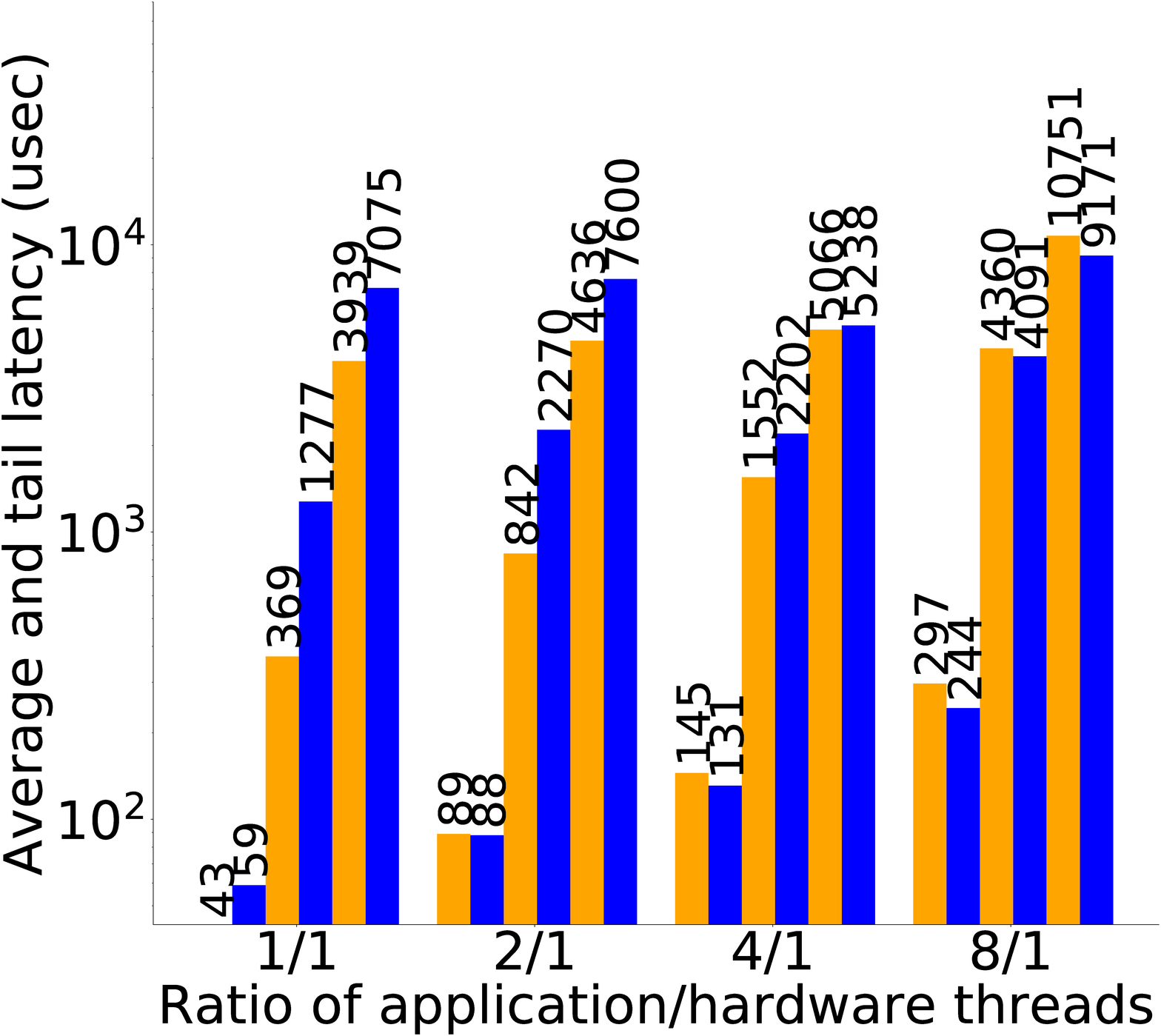}\label{subfig:load_a_avg_rdb}}~\subfigure[RocksDB, Run C]{\includegraphics[width=.49\linewidth]{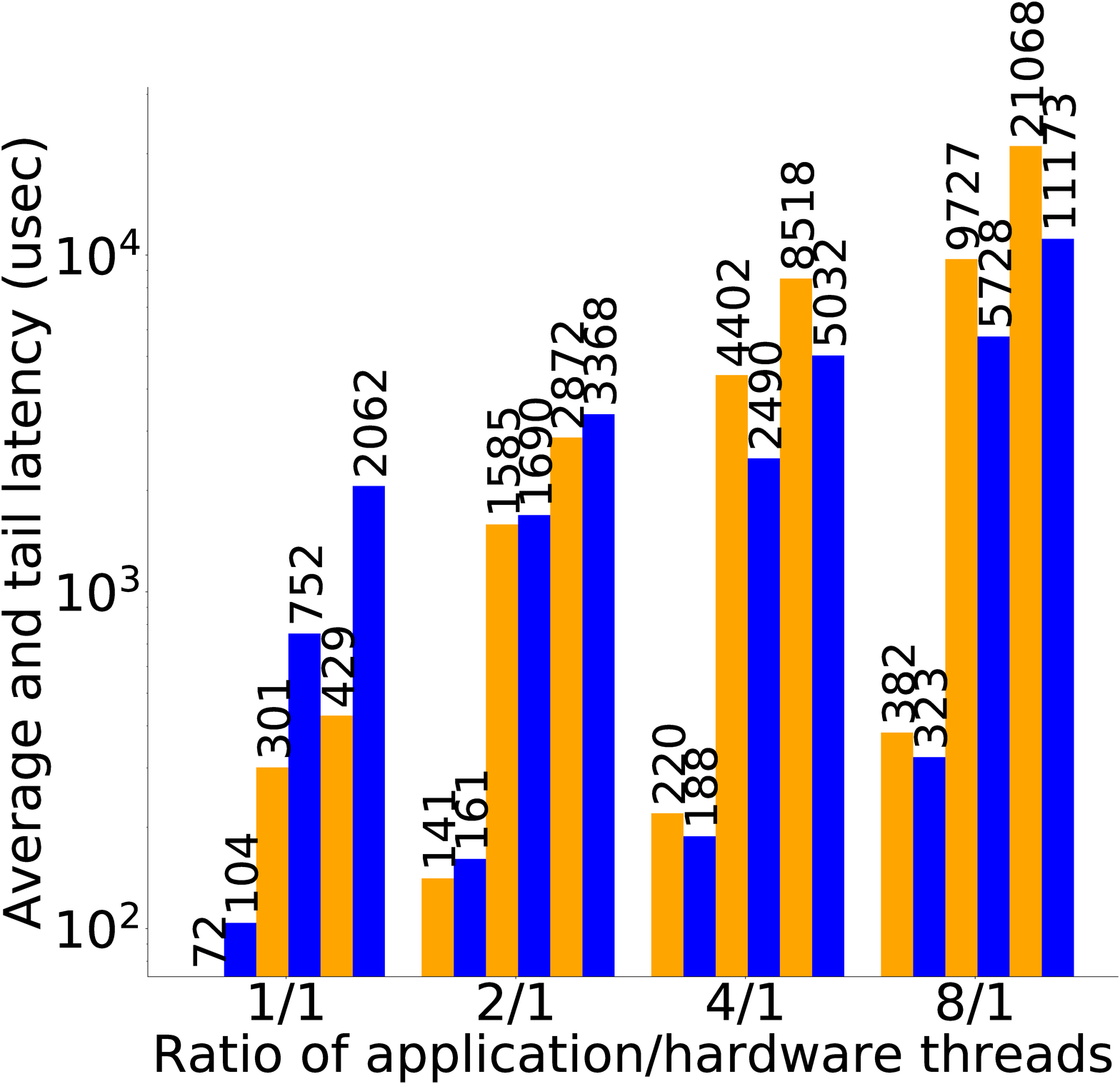}\label{subfig:run_c_avg_rdb}}
\caption{Average and tail latency ($\mu$s) for Load A and Run C with Kreon (top) and RocksDB (bottom).}
\label{fig:tail_with_avg}
\end{figure}

To capture how tail latency deteriorates as load increases, we
increase the number of application threads per hardware threads on 
each server from 1-to-1 up to 8-to-1. We also examine lower and higher 
loads, but we find this range to be representative. Figure~\ref{fig:tail_with_avg} 
shows the average and tail latency for S1 and S4 and two workloads, Load A and Run C, for both
Kreon and RocksDB. We use only S1 and S4 because these two server types 
exhibit the largest difference in performance. For Kreon we focus 
on in-memory KV store performance, where the server and CPU type 
has the highest impact. For this reason, we use 3M keys in Load A for 
both S1 and S4, where they fit in memory in both servers. Then, for Run C 
we run a larger number of get operations to extend the execution time  
to 5 minutes on each server and obtain reliable 99.9\% tail 
latency measurements.

In Figures~\ref{subfig:load_a_avg} and ~\ref{subfig:run_c_avg}, we see 
that average response time differs as follows: S4 has slightly lower 
average latency compared to S1 for Load A. More specifically, S4 average
response time is 17 $\mu$s, compared to 23 $\mu$s. In Run C however, 
S1 and S4 have almost the same average latency 8$\mu$s for S1 vs. 9$\mu$s
for S4. We notice that as load increases, tail latency deteriorates 
significantly for both workloads and both servers, 
to hundreds of times compared to average latency at high load. In Load A, 
tail latency becomes up to 107x worse (S4, 8-to-1 99.9\%), whereas in Run C, 
tail latency becomes up to 626x worse (S1, 4-to-1, 99.9\%) compared to average 
latency in the same run.  We also observe that 
generally, tail latency deteriorates in a similar manner on both servers, without
one of the two servers exhibiting worse behavior compared to the
other.

For RocksDB we run two different experiments. In the first experiment, 
we examine the performance of RocksDB with I/O traffic. 
For both Load A and Run C, we use the same dataset size as in the 
multi-threaded experiment (5M keys for S1 and 20M keys for S4). As 
we use a 4x larger dataset in S4 we also use a 4x larger 
user-space block cache (2GB for S1 and 8GB for S4) in combination 
with direct I/O to bypass the Linux kernel buffer cache. For Run C, 
we use a number of operations (\textit{gets}) required for 5 
minute run for all cases.
Figure~\ref{subfig:load_a_avg_rdb} shows for Load A that in all cases 
(except 1-to-1) S4 has slightly better average latency compared to S1. 
In the case of 1-to-1, S4 has an average latency of 59~$\mu$s compared 
to 43~$\mu$s for S1.
Figure~\ref{subfig:run_c_avg_rdb} shows for Run C that when we have 
low load (1-to-1 and 2-to-1) S1 has better latency time compared to S4. 
In case of high load (4-to-1 and 8-to-1) S4 becomes better compared
to S1.
We observe that as the load is increased further, tail 
latency also increases for both S1 and S4, for both workloads 
(Load A and Run C). Specifically, in Load A, tail latency 
becomes up to 120x worse (S4, 1-to-1, 99.9\%), whereas in Run C, 
tail latency increases up to 55x (S1, 8-to-1, 99.9\%) compared to average 
latency.

Next we examine how the size of the block cache 
affects the average and tail latency. We provide results only for 
Run C, as for loads RocksDB bypasses the block cache. We use the same
setup as in the first experiment, and only show the case with high 
load (8-to-1). We keep the ratio of dataset to block cache size 
the same for both S1 and S4. We start with the 
block cache disabled and then increase its size as follows: 512MB/2GB, 
1GB/4GB and 2GB/8GB, for S1/S4 respectively. 
\begin{figure}[t]
  \centering
  \includegraphics[width=1\linewidth]{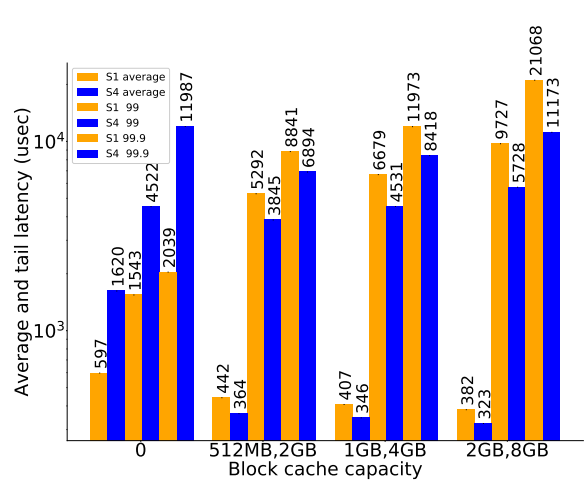}
  \caption{Average and tail latency ($\mu$s) for Run C in
    RocksDB with different I/O block cache sizes.}
  \label{fig:tail_with_avg_rdb_diff}
\end{figure}

%
Figure~\ref{fig:tail_with_avg_rdb_diff} shows the results of this 
experiment. When there is no block cache, S1 is much
better and up to 5.9x compared to S4 for both average and tail 
latencies. In this case we observe that the average disk queue 
depth is 254 for S4 and 10 for S1. 
As S4 has faster CPU compared to S1, it also processes requests and sends
them to the device at a higher rate. This results in more pressure 
to the device and higher average and tail latencies.

Finally, as we increase the size of the block cache, both S1 and S4 achieve 
better average and worse tail latencies. Increasing the size of the block 
cache means that a larger part of the dataset fits in the cache, 
resulting in lower average latency (higher hit ratio). However,
a larger cache also results in slower hits, misses and evictions, which 
increase the tail latency.
The average latencies of S1 are slightly worse compared to S4, with the 
a difference of 1.21x when the block cache ratio is 512MB/2GB (S1/S4).
In this case, the average latency of S1 is 442 $\mu$s vs. 364 $\mu$s 
for S4.
By increasing the block cache size, S1 always has higher tail latency 
compared to S4, up to 1.88x for a block cache ratio of 2GB/8GB for 99.9\%.

\section{Conclusions}
\label{sec:conclusion}


Persistent KV stores are an important component for modern
software stacks in the data center. In this work we 
provide an extensive power/performance evaluation for persistent KV stores. 
We examine how the processor micro-architecture 
and memory hierarchy affect data serving systems. We use four server 
types and two different KV stores (Kreon~\cite{kreon} and RocksDB) 
to measure power efficiency and absolute performance. 

A microserver (S1) results in 1.6-3.6x better power efficiency
compared to an x86 server with the same fabrication technology
(S2). S1 is up to 1.87x more power efficient compared to S4, a more
powerful server of newer process technology (22nm vs. 40nm). Although
all processors have similar CPU clocks, servers with more cores result
in higher performance. S4, with 2x more physical cores from S1 and 
hyper-threading enabled, achieves up to 5.3x more operations per second than S1.
All of these come with small impact in
tail latency.  Our analysis shows that architectural features such
as aggressive branch predictors, large caches, and hyper-threading do
not provide significant benefits in performance. The most significant
performance benefit comes from better memory throughput.

\commentout{
We perform a cost analysis based on energy cost, which shows that
S1 has 1.1-2.7x lower energy cost. If in addition we include
server equipment cost and a depreciation period of 3 to 5 years
(server lifetime), then total cost efficiency depends on server
purchase price. For microservers, such as S1 to be more cost-effective
they need to have a purchase price several times lower, and typically
around or more than 3x, than higher end servers, such as S4.

In summary, the most appropriate solution for KV stores is
microservers with large numbers of cores, relatively simple branch
prediction, small caches, no hyper-threading, and large memory
throughput; however, we need to consider from an economic perspective that microservers 
will also have a significantly lower purchase price compared to high-end servers.
}

\section*{Acknowledgements}
We thankfully acknowledge the support of the European Commission
through the H2020 project EVOLVE (GA825061). We would like also
to thank Michalis Flouris for his help with access to the ARM server. 
Anastasios Papagiannis is also supported by the Facebook Graduate 
Fellowship.

\bibliographystyle{plain}
\bibliography{ms}
\end{document}